\pdfoutput=1 
\documentclass[a4paper, 11pt]{article}
\usepackage[dvipsnames]{xcolor}

\usepackage{jheppub} 

\makeatletter  
\gdef\@fpheader{}  
\makeatother 

\usepackage[T1]{fontenc} 
\usepackage{tangocolors}
\usepackage{tikz}
\usepackage{tikz-3dplot}
\usetikzlibrary{arrows, decorations,backgrounds, patterns}
\usetikzlibrary{decorations.pathreplacing ,decorations.markings}
\usepackage{amssymb}
\usetikzlibrary{decorations.pathmorphing,backgrounds,shapes,arrows,shadows}
\usetikzlibrary{patterns.meta}
\usetikzlibrary{patterns,decorations.pathmorphing}
\tikzset{
    snake it/.style={decorate, decoration=snake}
}
\usepackage{pgfplots}
\pgfplotsset{compat=1.11}
\usepgfplotslibrary{fillbetween}
\usetikzlibrary{intersections}
\pgfdeclarelayer{bg}
\pgfsetlayers{bg,main}
\tikzset{zigzag/.style={decorate,decoration=zigzag}}
\tikzset{snake it/.style={decorate, decoration=snake}}
\makeatletter
\def\@hex@@Hex#1%
 {\if a#1A\else \if b#1B\else \if c#1C\else \if d#1D\else
  \if e#1E\else \if f#1F\else #1\fi\fi\fi\fi\fi\fi \@hex@Hex}
\makeatother

\usepackage[all]{xy}
\usepackage[percent]{overpic}
\usepackage{slashed}
\usepackage{wrapfig}
\usepackage{tabu}
\usepackage{diagbox}
\usepackage{mathrsfs,amsmath,amssymb,amsthm,amsfonts,tikz,graphicx,accents,hyperref, color}
\usepackage{dsfont,epiolmec, latexsym, stmaryrd, comment}
\usepackage{slashed,ccaption}
\usepackage{mathrsfs, calligra}
\usepackage{leftidx}
\usepackage{import}
\usepackage{multirow}
\usepackage{amsfonts}
\usepackage{pifont}
\usepackage{tabularx}
\usepackage{cancel}
\usepackage[utf8]{inputenc}
\usetikzlibrary{intersections,calc}
\usepackage{ifthen}
\usepackage{amsmath}
\usepackage{physics}
\usepackage{cancel}
\usepackage{relsize}
\usepackage{caption} 
\usepackage{subcaption}

\usetikzlibrary{patterns.meta}
\usepackage{array}
\usepackage{bm}

\usepackage{stmaryrd}

\hypersetup{ linktoc=all,
    colorlinks, linkcolor={palatinateblue},
    citecolor={brightpink}, urlcolor={amaranth}
}

\graphicspath{{Images/}}

\renewcommand{\d}[1]{\ensuremath{\operatorname{d}\!{#1}}}

\def\sideremark#1{\ifvmode\leavevmode\fi\vadjust{\vbox to0pt{\vss
 \hbox to 0pt{\hskip\hsize\hskip1em
 \vbox{\hsize2cm\tiny\raggedright\pretolerance10000
 \noindent #1\hfill}\hss}\vbox to8pt{\vfil}\vss}}}%
\DeclareSymbolFont{extraup}{U}{zavm}{m}{n}
\DeclareMathSymbol{\varheart}{\mathalpha}{extraup}{86}
\DeclareMathSymbol{\vardiamond}{\mathalpha}{extraup}{87}
\makeatletter
\renewcommand*{\@fnsymbol}[1]{\ensuremath{\ifcase#1\or \clubsuit \or \vardiamond \or \varheart\or
    \spadesuit\or \mathparagraph\or \|\or **\or \dagger\dagger
    \or \ddagger\ddagger \else\@ctrerr\fi}}
\makeatother

\definecolor{rosy}{RGB}{230,235,252}
\definecolor{myframetitle}{RGB}{90,89,170}
\definecolor{myblocktitle}{RGB}{140,185,249}
\definecolor{mytitle}{RGB}{10,80,26}

\definecolor{darkgreen}{RGB}{27,130,45}
\definecolor{darkblue}{rgb}{0,0,0.3}
\definecolor{darkred}{rgb}{0.7,0,0}

\definecolor{light gray}{RGB}{220,220,220}
\definecolor{dark purple}{RGB}{108,0,217}
\definecolor{pink}{RGB}{190,20,100}
\definecolor{orang}{RGB}{193,63,0}
\definecolor{green}{RGB}{11,98,17}
\definecolor{darkpink}{RGB}{153,0,76}
\definecolor{bluegreen}{RGB}{0,102,102}
\definecolor{greenlagan}{RGB}{0,102,0}
\definecolor{redgreen}{RGB}{102,102,0}
\definecolor{Redgreen}{RGB}{153,76,0}
\definecolor{vividviolet}{rgb}{0.62, 0.0, 1.0}
\definecolor{amaranth}{rgb}{0.9, 0.17, 0.31}
\definecolor{palatinateblue}{rgb}{0.15, 0.23, 0.89}
\definecolor{brightpink}{rgb}{1.0, 0.0, 0.5}
\definecolor{cornflowerblue}{rgb}{0.39, 0.58, 0.93}
\definecolor{deepcarminepink}{rgb}{0.94, 0.19, 0.22}
\definecolor{radicalred}{rgb}{1.0, 0.21, 0.37}
\definecolor{darkmagenta}{rgb}{0.67, 0, 0.67}

\usepackage[most]{tcolorbox}

\tcbset{highlight math style={left=02mm,right=02mm,top=02mm,bottom=02mm}} 
\usepackage{empheq}

\newcommand\inbox[1]{\tcbset{fonttitle=\scriptsize} \tcboxmath[colback=white,colframe=black!70]{#1}}
\newcommand{\Ottbar}{{\mathcal{T}\hspace*{-1mm}{\mathcal{T}}}}
\newcommand{\ttbar}{{\mathrm{T}\hspace*{-1mm}{\mathrm{T}}}}

\newcommand{\ttbarb}{{\mathrm{t}\hspace*{-1mm}{\mathrm{t}}}}

\DeclareFontFamily{OT1}{rsfs}{}

\DeclareFontShape{OT1}{rsfs}{m}{n}{ <-7> rsfs5 <7-10> rsfs7 <10->rsfs10}{} 

\DeclareMathAlphabet{\mycal}{OT1}{rsfs}{m}{n}

\newcommand{\be}{\begin{equation}}
\newcommand{\ee}{\end{equation}}
\newcommand{\bea}{\begin{eqnarray}}
\newcommand{\eea}{\end{eqnarray}}
\makeatletter \@addtoreset{equation}{section}

\usepackage{stackengine}

\begin{document}


\newcommand{\mytitle}{\begin{center}{\LARGE{GR from RG, $2d$ Example:}\\ \Large{JT-Gravity Induced from Renormalization Group Flow}}
\end{center}}

\title{{\mytitle}}
\author[a]{M.M.~Sheikh-Jabbari}
\author[b,a]{, V.~Taghiloo}

\affiliation{$^a$ Department of Physics, Institute for Advanced Studies in Basic Sciences (IASBS),\\ 
P.O. Box 45137-66731, Zanjan, Iran}
\affiliation{$^b$ School of Physics, Institute for Research in Fundamental
Sciences (IPM),\\ P.O.Box 19395-5531, Tehran, Iran}
\emailAdd{
jabbari@theory.ipm.ac.ir, v.taghiloo@iasbs.ac.ir}

\abstract{We demonstrate how  the two-dimensional gravity emerges within ``GR from RG'' program initiated in \cite{Adami:2025pqr, Sheikh-Jabbari:2026uol}. To achieve this, we consider a generic 2d CFT with a 3d holographic description, which we assume to be well-described by pure Einstein-AdS$_3$ gravity in the bulk. We study the holographic RG flow for the 2d CFT action and show that the renormalization group (RG) corrected action at an arbitrary energy scale contains a 2d scalar-tensor gravity theory. In the simplest case, the flow induces Jackiw-Teitelboim (JT) gravity, where the bulk radial lapse function seeds the dynamical dilaton field of the JT gravity. We show that the standard T$\bar{\text{T}}$  deformation of the 2d CFT is recovered as a special case in the Fefferman-Graham limit where the lapse is fixed. We further establish the robustness of the RG induced gravity picture by verifying its consistency under holographic renormalization and by generalizing the result to a one-parameter family of boundary conditions. Our results provide a first-principles derivation of the JT gravity at a finite cutoff as an intrinsic manifestation of the holographic RG flow in a non-Fefferman-Graham gauge. }

\maketitle


\section{Introduction}\label{sec:Intro}

\paragraph{Motivation.}
The nature of gravity remains one of the most profound puzzles in theoretical physics. While traditionally viewed as a fundamental force, an alternative paradigm suggests that gravity is an \textit{emergent} phenomenon—a coarse-grained, collective description of underlying non-gravitational microscopic degrees of freedom. This viewpoint, pioneered by Sakharov \cite{Sakharov:1967pk} and echoed in black hole thermodynamics \cite{Bekenstein:1973ur, Hawking:1975vcx} and the fluid/gravity correspondence \cite{Bhattacharyya:2007vjd, Rangamani:2009xk, Hubeny:2011hd}, has recently been revitalized through the ``GR from RG'' program initiated in \cite{Adami:2025pqr, Sheikh-Jabbari:2026uol}. In this framework, the Einstein-Hilbert action is not postulated as fundamental but is instead \textit{induced} by the renormalization group (RG) flow of a quantum field theory (QFT). The central idemechanism relies on the fact that every QFT possesses a universal coupling to its energy-momentum tensor \cite{Zamolodchikov:2004ce}; under holographic RG flow, deformations by operators quadratic in the stress tensor (such as $\text{T}\bar{\text{T}}$) \cite{Smirnov:2016lqw, Cavaglia:2016oda} naturally map to gravitational dynamics via bulk constraints \cite{McGough:2016lol, Hartman:2018tkw, Taylor:2018xcy}.

In this work, we explore the emergence of two-dimensional (2d) gravity within this program. Studying 2d gravity is a worthwhile and highly instructive exercise for several reasons. First, 2d models like Jackiw-Teitelboim (JT) gravity \cite{Jackiw:1984je, Teitelboim:1983ux} serve as the essential laboratory for understanding the quantum structure of black hole horizons \cite{Almheiri:2014cka} and the holographic nature of nearly-AdS$_2$ spacetimes \cite{Maldacena:2016upp, Jensen:2016pah}. Second, while 2d Einstein gravity is topologically trivial, scalar-tensor theories provide a rich dynamical landscape. Finally, the lower dimensionality of the AdS$_3$/CFT$_2$ correspondence \cite{Maldacena:1997re, Brown:1986nw} allows for a level of analytical precision—often yielding exact, closed-form solutions—that is difficult to achieve in higher dimensions. By demonstrating how JT gravity emerges from a generic 2d CFT, we provide a robust lower-dimensional anchor for the GR from RG program.

\paragraph{Setting and analysis.}
We consider a $(1+2)$-dimensional bulk spacetime described by pure Einstein gravity with a negative cosmological constant. According to the holographic dictionary \cite{Gubser:1998bc, Witten:1998qj}, this bulk is dual to a 2d conformal field theory (CFT) residing on the asymptotic boundary. We employ a radial foliation of the bulk, where the radial coordinate $r$ is identified with the RG energy scale of the boundary theory \cite{Susskind:1998dq, Peet:1998wn, deBoer:1999tgo, Skenderis:1999nb, deBoer:2000cz, Skenderis:2002wp, Heemskerk:2010hk, Faulkner:2010jy}. Crucially, we work in a coordinate system where the radial lapse function $\Phi$ is kept as an arbitrary degree of freedom, rather than being rigidly fixed as in the standard Fefferman-Graham gauge \cite{Fefferman1985, fefferman2012ambient}.

Our analysis proceeds in several key steps. We first derive the holographic RG flow equation for the boundary action at a finite radial cutoff, showing that it is driven by the 2d T$\bar{\text{T}}$ operator. By invoking the bulk Hamiltonian constraint, we replace this quadratic stress-tensor combination with boundary geometric quantities, effectively transforming the RG flow of the field theory into the radial evolution of a gravitational action.\footnote{Related studies investigating the connection between stress-tensor deformations and emergent gravitational dynamics via alternative frameworks include \cite{Ran:2025xas, Li:2025lpa, Xie:2026kek}.} We then perform an exact integration of this flow equation to find the effective boundary action at an arbitrary energy scale.

The primary result of this derivation is the emergence of a JT-like gravity theory on the cutoff surface, where the bulk radial lapse $\Phi$ is identified as the physical source for the induced dilaton field $\Psi$. Furthermore, we demonstrate that the ``unfreezing'' of the boundary metric—the transition of the background from a static structure to a dynamical field—is a direct consequence of the \textit{RG flow of boundary conditions}: the rigid Dirichlet condition imposed at asymptotic infinity naturally evolves into a mixed Dirichlet-Neumann condition in the IR. We verify that this mechanism is consistent across both bare and holographically renormalized actions \cite{deHaro:2000vlm} and holds for a generalized family of boundary conditions.

\paragraph{Main Results.} The central result of this work is the exact, non-perturbative derivation of a two-dimensional gravity action from the holographic RG flow of a three-dimensional bulk. Specifically, we track the radial evolution of the boundary action from the UV conformal boundary to a finite-radius cutoff surface. We show that this evolution is governed by a deformation flow equation which, once integrated, yields an effective 2d scalar-tensor gravity theory on the cutoff surface. In the simplest case, this induced theory takes the form of Jackiw-Teitelboim gravity:
\begin{equation}
    S_{\text{induced}} \propto \int \d{}^2x \sqrt{-\gamma}\ \Psi \left(R[\gamma] - 2\Lambda(r)\right) , \nonumber
\end{equation}
where the dynamical dilaton field $\Psi$ is uniquely sourced by the bulk radial lapse function $\Phi$. 

We demonstrate that the standard $\text{T}\bar{\text{T}}$ deformation is a special case of this result, recovered precisely in the Fefferman-Graham limit where the lapse is held fixed. Furthermore, we prove that this mechanism is universal: for a generalized family of renormalized and non-renormalized boundary conditions, the radial flow always induces a JT-like dynamics, with the specific choice of boundary condition merely selecting a unique trajectory in the space of effective couplings. Finally, we provide a rigorous consistency check by showing that the integrability of this boundary RG flow is exactly equivalent to the transverse tensor components of the bulk Einstein equations.

\paragraph{Organization of the paper.} 
In Section \ref{sec:setup-dynamics}, we establish the geometric foliation of the bulk and solve the AdS$_3$ Einstein equations perturbatively near the boundary. Section \ref{sec:RG-flow-action-BC} derives the holographic RG flow equation and investigates the radial evolution of boundary conditions. In Section \ref{sec:Int-bdry-action}, we present our main result: the exact integration of the flow equation that yields the induced JT action. Section \ref{sec:renormalization} repeats this derivation within the framework of holographic renormalization. We explore the universality of this induced gravity across a generalized family of boundary conditions in Section \ref{sec:other-bc}. Finally, we offer concluding remarks in Section \ref{sec:conc}. Technical identities are collected in Appendix \ref{sec:relations}.

\section{Geometric setup and asymptotic dynamics}\label{sec:setup-dynamics}
In this section, we establish the geometric and dynamical framework required for our analysis. We begin by introducing a radial foliation of the spacetime, which naturally decomposes the metric into boundary and transverse components. We then define the bulk gravitational theory and systematically solve its equations of motion near the asymptotic boundary. This setup provides the foundational variables and their asymptotic behaviors, which are essential for driving the holographic renormalization group flow in the subsequent sections.

\subsection{Geometric foliation}\label{sec:setup}

\paragraph{The spacetime arena.} We consider a $(1+2)$-dimensional spacetime $\mathcal{M}$ that is asymptotically anti-de Sitter (AdS), equipped with coordinates $x^\mu$ and a Lorentzian metric $g_{\mu\nu}$. To track the holographic evolution from the ultraviolet (UV) to the infrared (IR), we introduce a radial foliation of $\mathcal{M}$ by a family of timelike hypersurfaces $\Sigma_r$. Each slice is parameterized by a radial coordinate $r \in (r_\circ, \infty)$, where $r_\circ \geq 0$ represents the innermost extent of the foliation. 

\begin{figure}[t]
\centering
\begin{tikzpicture}[scale=1.3]
    \def\ROut{2.2}     
    \def\HOut{4.0}     
    \def\RIn{1.1}      
    \def\YEll{0.4}     
    \def\YEllIn{0.25}  
    
    \draw[blue!40, dashed, thick] plot[variable=\t, domain=0:180, samples=50] ({\ROut*cos(\t)}, {\YEll*sin(\t)});
    \draw[blue!40, dashed, thick] plot[variable=\t, domain=0:180, samples=50] ({\ROut*cos(\t)}, {-\HOut + \YEll*sin(\t)});
    
    \shade[left color=red!40!gray!30, right color=red!40!gray!30, middle color=red!5!white, draw=darkred!80, thick, opacity=0.85]
        (-\RIn, 0) -- (-\RIn, -\HOut)
        -- plot[variable=\t, domain=180:360, samples=40] ({\RIn*cos(\t)}, {-\HOut + \YEllIn*sin(\t)})
        -- (\RIn, -\HOut) -- (\RIn, 0)
        -- plot[variable=\t, domain=360:180, samples=40] ({\RIn*cos(\t)}, {\YEllIn*sin(\t)})
        -- cycle;
        
    \filldraw[fill=gray!20, draw=darkred!80, thick] plot[variable=\t, domain=0:360, samples=60] ({\RIn*cos(\t)}, {\YEllIn*sin(\t)});
    
    \foreach \h in {-0.8, -1.6, -2.4, -3.2} {
        \draw[darkred!50, thin, opacity=0.7] plot[variable=\t, domain=180:360, samples=50] 
            ({\RIn*cos(\t)}, {\h + \YEllIn*sin(\t)});
    }
    
    \draw[blue!70, thick] (-\ROut, 0) -- (-\ROut, -\HOut);
    \draw[blue!70, thick] (\ROut, 0) -- (\ROut, -\HOut);
    \draw[blue!70, thick] plot[variable=\t, domain=180:360, samples=50] ({\ROut*cos(\t)}, {-\HOut + \YEll*sin(\t)});
    \draw[blue!70, thick] plot[variable=\t, domain=0:360, samples=60] ({\ROut*cos(\t)}, {\YEll*sin(\t)});
    
    \draw[->, >=stealth, thick, black!80] (0, 0.8) -- (\ROut, 0.8) node[midway, above] {$r \to \infty$};
    \draw[dashed, black!50] (0,0) -- (0, 0.8);
    \draw[dashed, black!50] (\ROut,0) -- (\ROut, 0.8);
    
    \node[blue!80!black, right, font=\large] at (\ROut + 0.1, -\HOut/2) {$\Sigma$};
    \node[darkred!80, right, font=\large] at (\RIn + 0.2, -\HOut/2 - 0.5) {$\Sigma_c$};
    \node[black, font=\Large] at (0, -\HOut/2) {$\mathcal{M}_c$};
    
\end{tikzpicture}
\caption{\footnotesize{A radial foliation of an asymptotically AdS$_3$ spacetime. The outer transparent cylinder represents the rigid conformal boundary $\Sigma$ at $r \to \infty$. The inner shaded subregion $\mathcal{M}_c$ is bounded by a timelike cutoff hypersurface $\Sigma_c$ at a finite radius $r = r_c$.}} \label{fig:ADS-timelike}
\end{figure}
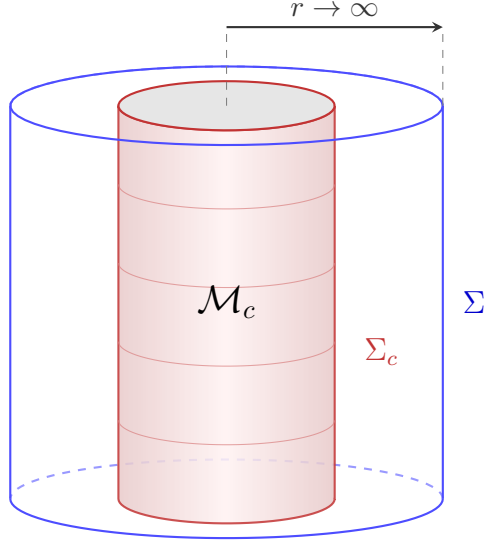

This foliation naturally induces a decomposition of the spacetime coordinates as $x^\mu = (x^a, r)$, where $a \in \{0, 1\}$. The coordinates $x^a$ chart the intrinsic geometry of the hypersurfaces $\Sigma_r$, while $r$ serves as the transverse coordinate. We identify the conformal boundary of the AdS spacetime with the limit $r \to \infty$, denoting this asymptotic boundary as $\Sigma := \Sigma_{r=\infty}$. To study the effective physics at an arbitrary energy scale, we consider a bulk subregion $\mathcal{M}_c$ enclosed by a finite radial cutoff at $r = r_c$, bounded by the surface $\Sigma_c := \Sigma_{r_c}$ (see Figure \ref{fig:ADS-timelike}). More generally, we denote by $\mathcal{M}_r$ the bulk region capped at a generic fixed radius $r$, with $\Sigma_r$ acting as its timelike boundary.

\paragraph{Metric decomposition and gauge choice.} 
The radial foliation naturally lends itself to an ADM-like decomposition \cite{Arnowitt:1962hi} of the spacetime metric. The generic line element assumes the form
\begin{equation}\label{metric}
    \d s^2 = \frac{\ell^2}{r^2} \Phi^2 \d r^2 +  h_{ab} (\d x^a + U^a \d r)(\d x^b + U^b \d r)\,,\qquad h_{ab}:=\frac{r^2}{\ell^2} \gamma_{ab}\, ,
\end{equation}
where $\Phi$ is the rescaled radial lapse function, $U^a$ represents the shift vector tangential to $\Sigma_r$, and $\gamma_{ab}$ is the conformal induced metric on each radial slice. To guarantee that the spacetime is asymptotically AdS, we impose the following standard fall-off conditions in the large $r$ limit:
\begin{equation}\label{Asymp-AdS-conditions}
    \Phi \sim \mathcal{O}(1)\, , \qquad \gamma_{ab} \sim \mathcal{O}(1)\, , \qquad U^{a} \sim \mathcal{O}(r^{-2})\, .
\end{equation}
It is important to note that the metric \eqref{metric} generally involves six free functions, and we have not yet fixed the 3d diffeomorphisms. A standard choice in holographic literature is the Fefferman-Graham (FG) gauge \cite{Fefferman1985, fefferman2012ambient}, which rigidly sets $\Phi=1$ and $U^a=0$, leaving only the three degrees of freedom encoded in the constant-$r$ metric $\gamma_{ab}$. 

However, \textit{for the purposes of our analysis, it is strictly crucial not to fix the radial lapse function $\Phi$}. As we will demonstrate, retaining $\Phi$ as an arbitrary degree of freedom is the physical mechanism that eventually seeds the dilaton field of the induced 2d gravity theory. In contrast, the shift vector $U^a$ does not play a dynamically generative role in this specific setup and may be conveniently set to zero later. Nonetheless, we keep $U^a$ unfixed for the time being to maintain full generality in our geometric definitions.

\paragraph{Geometric variables.}
The unit normal one-form $s$ to the hypersurfaces $\Sigma_r$, together with the corresponding projection tensor $h_{\mu\nu}$, are given by
\begin{equation}
    s = s_\mu \d x^\mu = \frac{\ell}{r}\Phi \d r\, , \qquad h_{\mu\nu} := g_{\mu\nu} - s_\mu s_\nu \, .
\end{equation}
Here, $h_{\mu\nu}$ restricts the spacetime metric onto the tangent space of the hypersurface orthogonal to $s^\mu$. We emphasize that henceforth, all lowercase Latin indices ($a,b,\dots$) on the boundary are exclusively raised and lowered using the conformal metric $\gamma_{ab}$.

The intrinsic dynamics of the foliation are governed by the extrinsic curvature of $\Sigma_r$, which encodes how each hypersurface is embedded within the bulk $\mathcal{M}$. It is defined via the Lie derivative along the normal vector \cite{Wald:1984rg, Misner:1973prb}:
\begin{equation}\label{Kab-def}
    K_{\alpha \beta} = \frac{1}{2} h^\mu{}_\alpha\, h^\nu{}_\beta\, \mathcal{L}_s h_{\mu\nu}\, . 
\end{equation}
To evaluate this efficiently, it is convenient to introduce the differential operator $\mathcal{D}_r := \partial_r - \mathcal{L}_U$, which incorporates both the strict radial derivative and the spatial shift contributions. Using this operator, the spatial components of the extrinsic curvature and its trace can be elegantly expressed in terms of the conformal metric $\gamma_{ab}$ as:
\begin{equation}\label{Kab}
    K_{ab} 
    = \frac{r^2}{\ell^3\, \Phi} \gamma_{ab} + \frac{r^3}{2\ell^3\, \Phi} \mathcal{D}_r \gamma_{ab}\, , \qquad  K 
    = \frac{2}{\ell\, \Phi} + \frac{r}{\ell\, \Phi} \mathcal{D}_r (\ln \sqrt{-\gamma})\, .
\end{equation}

\subsection{\texorpdfstring{AdS$_{3}$}{AdS3} gravity}\label{sec:AdS-grvity}

\paragraph{Action and variational principle.}
The dynamics of the bulk geometry  governed by the three-dimensional Einstein-Hilbert action with a negative cosmological constant. When formulated to be compatible with a Dirichlet boundary condition on the radial slice $\Sigma_r$, the action takes the form:
\begin{equation}\label{action-AdS}
    S^{\text{\tiny{D}}}_{\text{\tiny{bulk}}}[\mathcal{M}_r] = \frac{1}{2\kappa_3} \int_{\mathcal{M}_r} \d{}^{3}x\, \sqrt{-g} \left(\mathscr{R} + \frac{2}{\ell^2}\right) + \frac{r^2}{\ell^2 \kappa_3} \int_{\Sigma_r}\d{}^{2}x\, \sqrt{-\gamma}\, K \, ,
\end{equation} 
where \( \kappa_3 = 8\pi G \) is the gravitational coupling, and $\mathscr{R}$ is the bulk Ricci scalar. The boundary integral in \eqref{action-AdS} is the standard Gibbons-Hawking-York (GHY) term \cite{PhysRevD.15.2752, PhysRevLett.28.1082}, which is necessary to render the Dirichlet variational principle well-posed.

Varying the action \eqref{action-AdS} yields the bulk vacuum Einstein field equations:
\begin{equation}\label{eom-AdS}
    \mathscr{R}_{\mu\nu}-\frac{1}{2}\mathscr{R}\, g_{\mu\nu} -\frac{1}{\ell^2} g_{\mu\nu}=0\, , \qquad \mathscr{R}= -\frac{6}{\ell^2}\, .
\end{equation} 
Simultaneously, tracking the on-shell boundary terms from the variation provides the Dirichlet symplectic potential evaluated on $\Sigma_r$ \cite{Lee:1990nz, Iyer:1994ys, Sheikh-Jabbari:2025kjd, Sheikh-Jabbari:2026ouj}:    
\begin{equation}\label{sym-pot}
    \Theta_{\text{\tiny{D}}}(\Sigma_r) = -\frac{1}{2} \int_{\Sigma_r} \sqrt{-\gamma}\, T^{ab}\, \delta \gamma_{ab} +\text{corner terms}\, .
\end{equation}  

\paragraph{The Brown-York stress tensor.}
The momentum conjugate to the induced conformal metric $\gamma_{ab}$ in the symplectic potential \eqref{sym-pot} is the Brown-York energy-momentum tensor (BY-EMT) \cite{Brown:1992br}. It is defined entirely in terms of the boundary's extrinsic curvature as:
\begin{equation}\label{BY-EMT-Kab}
    T_{ab}=\frac{1}{\kappa_3}\left(K_{ab} - \frac{r^2}{\ell^2} K \gamma_{ab}\right)\, , \qquad T := \gamma^{ab} T_{ab} = - \frac{r^2}{\ell^2 \kappa_3} K\, .
\end{equation}  

\paragraph{Radial decomposition of field equations.}
To facilitate our holographic RG flow analysis, we  project the 3d equations of motion \eqref{eom-AdS} along and orthogonal to the foliation $\Sigma_r$. Using the metric decomposition \eqref{metric}, the Einstein equations  split into a set of constraints and radial evolution equations:
\begin{subequations}\label{EoM-GR-decompose}
    \begin{align}
       & {R} + \frac{2r^2}{\ell^4} + \kappa_3^{2} \frac{\ell^2}{r^2}\ttbar = 0 \, ,\label{EoM-ss}\\
       & {\nabla}_b {{T}}^{b}_{a}=0\, , \label{EoM-sa}\\
       & \frac{\kappa_3}{\Phi}\mathcal{D}_r {{T}}_{ab}
       -\frac{\ell^3 \kappa_3^2}{r^3} (T T_{ab} +2 \ttbar \gamma_{ab}) + \frac{\ell}{r} \left( \frac{{\nabla}_{a}{\nabla}_b \Phi}{\Phi}-\frac{\Box \Phi}{\Phi}\gamma_{ab} \right)-\frac{\ell}{2r}{R} \gamma_{ab}=0\, , \label{EoM-ab} \\
       & \mathcal{D}_r \gamma_{ab}  + \frac{2}{r} \gamma_{ab} - \frac{2\ell^3\kappa_3}{r^3} \Phi (T_{ab}-{T} \gamma_{ab}) = 0\, . \label{dr-h-T}
    \end{align}
\end{subequations}
Here, \( R \) denotes the intrinsic boundary Ricci scalar, and \( \nabla_{a} \) is the covariant derivative built from the Levi-Civita connection of \( \gamma_{ab} \). Note that \eqref{dr-h-T} is not a part of Einstein's equations, it is just the definition of BY-EMT \eqref{BY-EMT-Kab}, written for completion.   We have also introduced the T$\bar{\text{T}}$ operator \cite{Zamolodchikov:2004ce, Smirnov:2016lqw, Cavaglia:2016oda}, defined as the characteristic quadratic combination of the stress tensor:
\begin{equation}\label{def-TTbar}
    \ttbar := {{T}}^{ab}\, {{T}}_{ab} - {{T}}^2 \, .
\end{equation} 
Notice that in deriving the tensor evolution equation \eqref{EoM-ab}, we utilized the fact that the boundary is two-dimensional. This allows us to apply the Cayley–Hamilton identity to $T_{ab}$, yielding the exact algebraic relation:
\begin{equation}\label{identity}
    \begin{split}
        T_{a}^{c} T_{cb} = T T_{ab} + \frac{1}{2} \ttbar \gamma_{ab}\, . 
    \end{split}
\end{equation}

The set of equations \eqref{EoM-GR-decompose} establishes the dynamical backbone of our formulation. While similar decomposed equations were explored in \cite{Sheikh-Jabbari:2025kjd, Sheikh-Jabbari:2026ouj}, our expressions generalize those results by accommodating a completely arbitrary radial lapse function $\Phi$, which is an essential ingredient of our analysis, as we will see in the following sections. As expected from the standard ADM formalism \cite{Arnowitt:1962hi} (see \cite{Gourgoulhon:2007ue} for a review), the Hamiltonian \eqref{EoM-ss} and momentum \eqref{EoM-sa} constraints are purely intrinsic to the hypersurface and remain entirely independent of the choices made for the lapse and shift functions.

\subsection{Asymptotic solution space}\label{sec:asymptotic-solution-space}

\paragraph{Perturbative expansion near the boundary.}
To understand the UV behavior of the theory and prepare the necessary boundary data for our RG flow equations, we solve the equations of motion \eqref{EoM-GR-decompose} perturbatively near the AdS boundary ($r \to \infty$)  \cite{deHaro:2000vlm, Skenderis:2002wp}. For this asymptotic analysis, we fix the spatial shift vector to $U^{a}=0$, but critically, we leave the radial lapse $\Phi$ arbitrary. Compatible with the asymptotic AdS conditions \eqref{Asymp-AdS-conditions}, we introduce the following fall-off ansatz for the metric components:
\begin{equation}\label{fall-off}
    \begin{split}
         & \Phi(r, x^a) = 1 + \frac{\ell^2}{r^2} \Phi_1(x^a) + \frac{\ell^4}{r^4} \Phi_2(x^a) + \cdots\, , \\
         & \gamma_{ab}(r, x^a) = q_{ab}(x^a) + \frac{\ell^2}{r^2} C_{ab}(x^a) + \frac{\ell^4}{r^4} D_{ab}(x^a) + \cdots \, .
    \end{split}
\end{equation}
Here, $q_{ab}(x^a)$ is the arbitrary induced metric on the AdS conformal boundary. From this ansatz, it is straightforward to verify that the inverse conformal metric expands as
\begin{equation}
   {\gamma}^{ab} = q^{ab} - \frac{\ell^2}{r^2} C^{ab} - \frac{\ell^4}{r^4} \left( D^{ab} - C^{ac}C^{b}_c \right)+ \cdots \, .
\end{equation}

\paragraph{Solving the radial evolution.} Substituting these expansions into the radial  equations \eqref{EoM-GR-decompose}, we can determine the  metric coefficients $C_{ab}$ and $D_{ab}$ perturbatively, order by order in $1/r$. The result is:
\begin{equation}
   \begin{split}
       & C_{ab} = \frac{12\pi \ell^2}{c}t_{ab} - \left( \Phi_1 + \frac{\ell^2}{2} R[q]  \right) q_{ab}\, , \\
       & D_{ab} = - \frac{3\pi \ell^4}{2c} R[q] t_{ab} - \frac{\ell^2}{4} D_{a} D_b \Phi_1 + \frac{1}{2}\left( -\Phi_2 + \Phi_1^2 + \frac{\ell^4}{8} R[q]^2 + \frac{36\pi^2 \ell^4}{c^2} \ttbarb \right)q_{ab}\, .
   \end{split}
\end{equation}
In these expressions, $t_{ab}$ acts as an integration constant from the perspective of radial integration; physically, it corresponds to the holographic energy-momentum tensor of the dual boundary UV theory \cite{Balasubramanian:1999re, deHaro:2000vlm}. The quantity $c$ is the celebrated Brown-Henneaux central charge \cite{Brown:1986nw}:
\begin{equation}
    c:=\frac{12\pi \ell}{\kappa_3}\, .
\end{equation}
Furthermore, the Hamiltonian and momentum constraints \eqref{EoM-ss} and \eqref{EoM-sa} evaluated at the boundary impose strict conservation and trace anomaly conditions on $t_{ab}$ \cite{Henningson:1998gx, Balasubramanian:1999re}:
\begin{equation}\label{asymp-const}
    D_{a}t^{ab} = 0\, , \qquad t := q^{ab} t_{ab} =\frac{c}{24\pi} R[q]\, .
\end{equation}
Here, $D_a$ denotes the covariant derivative compatible with the boundary metric $q_{ab}$, and $\ttbarb$ is the $\text{T}\bar{\text{T}}$ operator  at the AdS boundary:
\begin{equation}
    \ttbarb := t_{ab} t^{ab} - t^2 \, .
\end{equation}

\paragraph{Unconstrained boundary data.}
At this stage, the asymptotic solution space is parameterized entirely by the boundary data $q_{ab}$, its conjugate momentum $t^{ab}$, and an infinite tower of arbitrary scalar fields $\Phi_i$ ($i=1,2,\dots$). It is crucial to emphasize that in writing these expansions, we have not yet imposed any specific boundary conditions. For instance, if one restricts to the well-known family of Ba\~nados geometries \cite{Banados:1998gg, Sheikh-Jabbari:2025kjd}, one fixes $\Phi_i=0$ and restricts $q_{ab}$ to a flat metric to strictly preserve standard Dirichlet boundary conditions, leading to $R[q]=0$ and $t=0$. In our setup, however, we keep the boundary fields and lapse coefficients completely general.

\paragraph{Semi-on-shell asymptotic quantities.}
To integrate the boundary action along the RG flow in later sections, we will need the asymptotic expansions of several composite geometric scalars and tensors. Substituting the solved metric coefficients back into our definitions, we obtain the following ``semi-on-shell'' expansions (meaning the bulk radial equations are satisfied, but the asymptotic constraints \eqref{asymp-const} are left explicit):
\begin{equation}\label{asymp-grav-quantities}
    \begin{split}
        & \sqrt{-\gamma} = \sqrt{-q} \left[ 1 - \left(\frac{\ell}{r}\right)^2\left( \Phi_1 + \frac{\ell^2}{4} R[q]\right) - \frac{1}{2}\left(\frac{\ell}{r}\right)^4 \left( \Phi_2 - \Phi_1^2 +\frac{\ell^2}{4} \Box \Phi_1 + \frac{36 \pi^2 \ell^4}{c^2}\ttbarb \right) + \cdots \right]\, , \\
        & R[\gamma] = R[q] + \frac{\ell^2}{r^2} \left( R[q]\Phi_1  + \frac{\ell^2}{4} R[q]^2 + \Box \Phi_1\right) + \cdots\, , \\
        & T_{ab} = -\frac{c}{12\pi \ell^2} \frac{r^2}{\ell^2}q_{ab}  - 2t_{ab} + \frac{c}{12\pi \ell^2} \left( \Phi_1 + \frac{\ell^2}{2} R[q]\right)q_{ab} + \cdots\, , \\
        & T = -\frac{c}{6\pi \ell^2} \frac{r^2}{\ell^2} -\frac{c R[q]}{24\pi} - \frac{\ell^2}{r^2}\left( \frac{6\pi \ell^2}{c}t_{ab}t^{ab} + \frac{c R[q]}{24\pi}\Phi_1 + \frac{c}{24\pi} \Box \Phi_1 \right) + \cdots\, .
    \end{split}
\end{equation}
These expansions, driven primarily by the boundary fields and the generic lapse coefficients $\Phi_1$ and $\Phi_2$, constitute the necessary mathematical toolkit to evaluate the boundary action at spatial infinity.

\section{Holographic RG flow of the action and boundary conditions}\label{sec:RG-flow-action-BC}
With the geometric foliation and asymptotic solution space established, we are now equipped to study the radial evolution of the boundary theory. In the holographic framework, moving the cutoff surface $\Sigma_r$ from the asymptotic boundary into the bulk corresponds to integrating out high-energy degrees of freedom, realizing an exact RG flow in the dual theory \cite{Heemskerk:2010hk, Faulkner:2010jy}. In this section, we derive the explicit flow equation governing the boundary action, rigorously verify its consistency with the bulk dynamics, and demonstrate a profound physical consequence: a rigid Dirichlet boundary condition imposed in the UV inevitably evolves into a dynamically mixed boundary condition in the IR.

\subsection{Dirichlet deformation flow equation}\label{sec:deformation-flow-D}

\paragraph{Radial evolution via diffeomorphisms.}
We first compute the radial deformation of the on-shell bulk action subject to Dirichlet boundary conditions. To do this, we employ the ``freelance holography'' machinery developed in \cite{Parvizi:2025shq, Parvizi:2025wsg} (see also \cite{Taghiloo:2025oeu} for a pedagogical review). A core insight of this approach is that the radial evolution of the on-shell action can be captured by evaluating its variation under a diffeomorphism generated by the radial vector field $\xi = \partial_{r}$. 

The variation of the Einstein-Hilbert action, evaluated on-shell and compatible with Dirichlet boundary conditions, under an arbitrary diffeomorphism $\xi$ is given purely by a boundary term:
\begin{equation}
    \delta_{\xi} S^{\text{\tiny{D}}}_{\text{bulk}}[\mathcal{M}_r] \Big|_{\text{on-shell}} = -\frac{\ell^2}{2r^2} \int_{\Sigma_r}\d{}^{2}x\, \sqrt{-\gamma}\, {T}^{ab}\, \delta_{\xi} \left(\frac{r^2}{\ell^2}\gamma_{ab}\right)\, . 
\end{equation}

\paragraph{The holographic T$\bar{\text{T}}$ flow equation.}
By strictly choosing $\xi = \partial_{r}$, we extract the exact radial derivative of the on-shell action:
\begin{equation}
   \begin{split}
       \frac{\d{}}{\d{}r} S^{\text{\tiny{D}}}_{\text{bulk}}[\mathcal{M}_r] \Big|_{\text{on-shell}} & = -\frac{\ell^2}{2r^2} \int_{\Sigma_r}\d{}^{2}x\, \sqrt{-\gamma}\, {{T}}^{ab}\, \partial_{r} \left(\frac{r^2}{\ell^2}\gamma_{ab}\right) \\
       & = - \frac{\ell^2}{r^2} \int_{\Sigma_r}\d{}^{2}x\, \sqrt{-\gamma}\, {{T}}^{ab}\, \left[ \frac{\ell \Phi}{r}  K_{ab} + \frac{r^2}{\ell^2}\nabla_{a} U_{b} \right] \\
       & = - \frac{\ell^3 \kappa_3}{r^3}\int_{\Sigma_r}\d{}^{2}x\, \sqrt{-\gamma}\, \Phi\, {{T}}^{ab}\, ({T}_{ab} - {T} \gamma_{ab})  \\
       & = - \frac{\ell^3 \kappa_3}{r^3} \int_{\Sigma_r}\d{}^{2}x\, \sqrt{-\gamma}\, \Phi\, \ttbar \, .
   \end{split}
\end{equation}
Let us unpack this sequence. The second line follows directly from substituting the definition of the extrinsic curvature \eqref{Kab}. The third line is achieved by replacing $K_{ab}$ with the Brown-York stress tensor via \eqref{BY-EMT-Kab} and integrating the shift vector term by parts, which vanishes due to the strict conservation of the stress tensor \eqref{EoM-sa} (${\nabla}_{a} {T}^{ab} = 0$). The final line elegantly compacts the expression using the standard definition of the $\text{T}\bar{\text{T}}$ operator \eqref{def-TTbar}.

Invoking the fundamental holographic dictionary at a finite radial cutoff, which equates the on-shell bulk action to the dual boundary action,
\begin{equation}
    S^{\text{\tiny{D}}}_{\text{bulk}}[\mathcal{M}_r] \Big|_{\text{on-shell}} = {S}^{\text{\tiny{D}}}_{\text{bdry}}[\Sigma_r]\, ,
\end{equation}
we obtain the celebrated deformation flow equation:
\begin{equation}\label{d-f-NR-D-TT}
   \inbox{ \frac{\d{}}{\d{}r} {S}^{\text{\tiny{D}}}_{\text{bdry}}[\Sigma_r] = - \frac{\ell^3 \kappa_3}{r^3}  \int_{\Sigma_r}\d{}^{2}x\, \sqrt{-\gamma}\, \Phi\, \ttbar \, .}
\end{equation}
This confirms that the radial evolution in pure Einstein-Hilbert gravity is intrinsically governed by a $\text{T}\bar{\text{T}}$-like deformation \cite{McGough:2016lol, Hartman:2018tkw, Taylor:2018xcy}.

\paragraph{Geometric formulation of the flow.}
While  \eqref{d-f-NR-D-TT} frames the flow in terms of the boundary fluid's stress tensor, it can be entirely recast into geometric variables. By substituting the Hamiltonian constraint \eqref{EoM-ss} to replace the $\ttbar$ operator with the intrinsic boundary curvature, we find an equivalent, purely gravitational representation of the flow:
\begin{equation}\label{d-f-NR-D-Gr}
  \inbox{  \frac{\d{}}{\d{}r} {S}^{\text{\tiny{D}}}_{\text{bdry}}[\Sigma_r] = \frac{c}{12\pi r} \int_{\Sigma_r}\d{}^{2}x\, \sqrt{-\gamma}\, \Phi\, \left(R[\gamma] + \frac{2r^2}{\ell^4} \right) \, .}
\end{equation}
Let us pause to examine the profound physical implications of the right-hand side of \eqref{d-f-NR-D-Gr}. Remarkably, the radial evolution of the boundary QFT is driven by a term that is exactly proportional to the Jackiw-Teitelboim gravity Lagrangian \cite{Jackiw:1984je, Teitelboim:1983ux}. In this expression, the bulk radial lapse function $\Phi$ naturally assumes the role of the 2d dynamical dilaton, coupling directly to the intrinsic boundary curvature $R[\gamma]$, while the $r^2/\ell^4$ term provides the corresponding cosmological constant. This is a striking realization of the ``GR from RG'' program: the generator of the renormalization group flow is itself a gravitational action. This structural appearance strongly suggests that the effective boundary action at any finite energy scale must inherently possess a JT gravity sector. In section \ref{sec:Int-bdry-action}, we will elevate this observation to a rigorous proof by introducing an exact JT-like ansatz for the finite-cutoff action and demonstrating that it non-perturbatively solves this very deformation flow equation.

\subsection{A consistency check: Derivation of bulk equations from boundary}\label{sec:derivation-bulk-eom}
As a robust consistency check of our framework—and a demonstration of the profound link between RG flow and bulk dynamics \cite{deBoer:1999tgo, deBoer:2000cz, Papadimitriou:2004ap}—we now show that the boundary deformation flow equation \eqref{d-f-NR-D-Gr} precisely encodes the bulk spatial equations of motion. If our formalism is consistent, the variation of the radial flow of the action must commute with the radial flow of its variation. That is, computing $\frac{\d{}}{\d{}r} \delta {S}^{\text{\tiny{D}}}_{\text{bdry}}[\Sigma_r]$ in two distinct ways must yield identical results. As we will see,  these two methods agree iff  the tensor Einstein equation \eqref{EoM-ab} holds.

\paragraph{Method 1: Variation of the flow equation.}
First, we directly vary the geometric flow equation \eqref{d-f-NR-D-Gr} with respect to the boundary metric. This trivially yields:
\begin{equation}\label{dr-delta-S-1}
    \frac{\d{}}{\d{}r} \delta {S}^{\text{\tiny{D}}}_{\text{bdry}}[\Sigma_r] = \frac{\ell}{r \kappa_3} \delta \int_{\Sigma_r}\d{}^{2}x\, \sqrt{-\gamma}\, \Phi\, \left(R[\gamma] + \frac{2r^2}{\ell^4} \right)  \, .
\end{equation}

\paragraph{Method 2: Radial flow of the variational principle.}
Alternatively, we can start from the definition of the boundary action's on-shell variation, driven by the Brown-York stress tensor:
\begin{equation}\label{dr-delta-S-2}
     \delta {S}^{\text{\tiny{D}}}_{\text{bdry}}[\Sigma_r] = -\frac{1}{2} \int_{\Sigma_r} \d{}^{2}x\, \sqrt{-\gamma}\, T^{ab} \delta \gamma_{ab}\, .
\end{equation}
Taking the radial derivative of this expression requires applying the chain rule to the integrand:
\begin{equation}\label{dr-delta-S-2-expanded}
         -\frac{1}{2}  \frac{\d{}}{\d{}r}\int_{\Sigma_r} \d{}^{2}x\, \sqrt{-\gamma}\, T^{ab} \delta \gamma_{ab}  = -\frac{1}{2}  \int_{\Sigma_r}\d{}^{2}x\, \Big[ {\cal D}_{r}(\sqrt{-\gamma}\, T^{ab}) \delta \gamma_{ab} + \sqrt{-\gamma}\, T^{ab} \mathcal{D}_r (\delta \gamma_{ab})  \Big]\, .
\end{equation}
To evaluate the last term, we use the commutation relation between the variation and the radial differential operator, $\mathcal{D}_r \delta \gamma_{ab} = \delta \mathcal{D}_r \gamma_{ab} + \mathcal{L}_{\delta U} \gamma_{ab}$. When multiplied by $\sqrt{-\gamma}\, T^{ab}$, the Lie derivative term $\mathcal{L}_{\delta U} \gamma_{ab}$ integrates to a corner term due to the conservation of the stress tensor \eqref{EoM-sa} ($\nabla_a T^{ab}=0$). Discarding this boundary term, we are left with evaluating $T^{ab} \delta {\cal D}_{r}\gamma_{ab}$.

Using the definition of the extrinsic curvature in terms of the stress tensor \eqref{dr-h-T} and applying the Cayley-Hamilton identity \eqref{identity}, we can rewrite this contraction as:
\begin{equation}
    T^{ab} \delta  {\cal D}_{r}\gamma_{ab} =  \left( -\frac{2T^{ab}}{r} + \frac{\ell^3 \kappa_3}{r^3}\Phi \ttbar \gamma^{ab} \right) \delta \gamma_{ab}  + \frac{2 \ell^3 \kappa_3}{r^3} \ttbar \delta \Phi + \frac{\ell^3 \kappa_3}{r^3} \Phi  \delta \ttbar\, .
\end{equation}
We can elegantly eliminate the $\delta \ttbar$ term by invoking the bulk Hamiltonian constraint \eqref{EoM-ss} and its variation:
 \begin{equation}
    \frac{\ell^3 \kappa_3}{r^3}\ttbar = -\frac{\ell}{\kappa_3 r} \left(R+\frac{2r^2}{\ell^4} \right)\,  \qquad \Longrightarrow \qquad  \frac{\ell^3 \kappa_3}{r^3}\delta \ttbar =  -\frac{\ell}{\kappa_3 r} \delta R\, .
 \end{equation}
Substituting this back, we can group the terms to explicitly reconstruct the variation of the JT-like action block:
\begin{equation}
    T^{ab} \delta  {\cal D}_{r}\gamma_{ab} =  -\frac{\ell}{\kappa_3 r} \Bigg\{ \frac{2\kappa_3}{\ell}T^{ab} \delta \gamma_{ab}  - \Phi \delta R + \frac{2}{\sqrt{-\gamma}} \delta\left[\sqrt{-\gamma}\, \Phi \left(R+\frac{2r^2}{\ell^4}\right)\right]\Bigg\}\, .
\end{equation}
Inserting this result into \eqref{dr-delta-S-2-expanded}, Method 2 yields:
\begin{equation}\label{Method-2-final}
    \begin{split}
        \frac{\d{}}{\d{}r} \delta {S}^{\text{\tiny{D}}}_{\text{bdry}}[\Sigma_r]=-\frac{1}{2}  \int_{\Sigma_r}\d{}^{2}x\, \Bigg\{ & \left[{\cal D}_{r}(\sqrt{-\gamma} T^{ab})-2\frac{\sqrt{-\gamma}}{r} T^{ab} \right]\delta \gamma_{ab} {+}\frac{\ell}{\kappa_3 r} \sqrt{-\gamma} \Phi \delta R \\
        &- \frac{2\ell}{\kappa_3 r} \delta\left[\sqrt{-\gamma}\, \Phi \left(R+\frac{2r^2}{\ell^4}\right)\right]\Bigg\} \, .
    \end{split}
\end{equation}

\paragraph{Synthesis and recovery of the equations of motion.}
We now demand that Method 1 \eqref{dr-delta-S-1} and Method 2 \eqref{Method-2-final} agree. Notice that the last term in \eqref{Method-2-final} is exactly equal to Method 1's expression. Therefore, for the two approaches to be strictly consistent, the remainder of the integral in \eqref{Method-2-final} must vanish identically:
\begin{equation}\label{consistency-condition-integral}
    \int_{\Sigma_r}\d{}^{2}x\, \Bigg\{ \left[{\cal D}_{r}(\sqrt{-\gamma} T^{ab})-2\frac{\sqrt{-\gamma}}{r} T^{ab} \right]\delta \gamma_{ab} {+}\frac{\ell}{\kappa_3 r} \sqrt{-\gamma} \Phi \delta R\Bigg\}= 0\, .
\end{equation}
To extract the local equations of motion, we utilize the standard variation of the Ricci scalar,
\begin{equation}
    \delta R = -\frac{1}{2}R \gamma^{ab} \delta \gamma_{ab} + \nabla^{a}\nabla^{b} \delta \gamma_{ab} - \gamma^{ab}\Box \delta \gamma_{ab}\, ,
\end{equation}
and perform integration by parts on the derivative terms to isolate $\delta \gamma_{ab}$. Dropping the resulting total derivatives, equation \eqref{consistency-condition-integral} demands:
\begin{equation}
    {\cal D}_{r}(\sqrt{-\gamma}\, T^{ab}) = \sqrt{-\gamma} \left[ \frac{2T^{ab}}{r} + \frac{\ell R}{2r\kappa_3}\Phi \gamma^{ab} - \frac{\ell}{r\kappa_3} (\nabla^{a}\nabla^{b}\Phi - \gamma^{ab} \Box \Phi)\right] \, .
\end{equation}
Remarkably, upon expanding the left-hand side using the definition of $T_{ab}$ \eqref{dr-h-T}, this is nothing other than the transverse tensor Einstein equation \eqref{EoM-ab}. We have thus successfully derived the spatial bulk dynamics strictly from the integrability of the boundary RG flow, relying solely on the Hamiltonian \eqref{EoM-ss} and momentum \eqref{EoM-sa} constraints.

\subsection{RG flow of the Dirichlet boundary condition}\label{sec:flow-D-bc}
It is a well-established property of Einstein gravity that it constitutes a well-defined initial-boundary value problem \cite{Friedrich:2000qv}. Once boundary data is specified at the asymptotic boundary $\Sigma$ alongside a chosen boundary condition, the bulk equations of motion uniquely determine the induced data and the effective boundary conditions on any other constant radial slice $\Sigma_r$. We now investigate this exact radial flow.

\paragraph{The asymptotic Dirichlet condition.}
We begin by imposing a strict Dirichlet boundary condition at the asymptotic conformal boundary $\Sigma$:
\begin{equation}\label{D-bc-Sigma}
 \text{Dirichlet b.c. at $\Sigma$}: \qquad   \delta q_{ab} = 0\, .
\end{equation}
While the induced metric $q_{ab}$ is absolutely fixed on $\Sigma$, its conjugate momentum $t_{ab}$ remains dynamical and undergoes arbitrary fluctuations. These fluctuations are only subject to the asymptotic Hamiltonian and momentum constraints \eqref{asymp-const}, which explicitly demand:
\begin{equation}
    \delta t = 0\, , \qquad D_{a} \delta t^{ab} = 0\, .
\end{equation}
Consequently, the UV dynamics of the dual theory are entirely governed by the conserved boundary stress tensor $t^{ab}$ and the arbitrary lapse fluctuations $\delta \Phi_i$. 

\paragraph{Flow of Dirichlet at finite distance.}
To see how this boundary condition evolves, we evaluate the variation of the induced metric at a finite radial distance using our asymptotic expansions \eqref{fall-off}. Tracking the variations yields:
\begin{equation}
    \begin{split}
        \delta \gamma_{ab}(r, x^a) =&\, \frac{\ell^2}{r^2} \left(\frac{12\pi \ell^2}{c} \delta t_{ab} -  \delta\Phi_1 q_{ab}\right) \\
        & +\frac{\ell^4}{r^4} \left(  - \frac{3\pi \ell^4}{2c} R[q] \delta t_{ab} - \frac{\ell^2}{4} D_{a} D_b \delta \Phi_1 + \frac{1}{2}\left( -\delta\Phi_2 + \delta\Phi_1^2 + \frac{36\pi^2 \ell^4}{c^2} \delta\ttbarb \right)q_{ab} \right) \\
        & + {\cal O}\big((\ell/r)^6\big)\, .
    \end{split}
\end{equation}
Crucially, although we imposed strict Dirichlet boundary conditions at infinity ($\delta q_{ab} = 0$), this condition is visibly not preserved under radial evolution. Instead, the fluctuations of the boundary stress tensor ($\delta t_{ab}$) and the lapse function ($\delta \Phi_1$) directly act as sources, driving metric fluctuations $\delta \gamma_{ab} \neq 0$ at any finite cutoff surface. 

\paragraph{Emergence of a mixed boundary condition.}
If the boundary condition is no longer Dirichlet, what is it? To answer this, we must similarly expand the variation of the Brown-York stress tensor at a finite distance using \eqref{asymp-grav-quantities}: 
\begin{equation}
    \delta T_{ab} = - 2 \delta t_{ab} + \frac{c}{12\pi \ell^2} \delta \Phi_1 q_{ab} +{\cal O}\big((\ell/r)^2\big)\, .
\end{equation}
Notice that the leading terms of $\delta \gamma_{ab}$ and $\delta T_{ab}$ are sourced by the exact same UV fluctuations ($\delta t_{ab}$ and $\delta \Phi_1$). By algebraically combining them to cancel these leading-order source terms, we uncover the true constraint operating at finite $r$:
\begin{equation}
    \delta \left(e^{(\Phi-1)/2} \gamma_{ab}(r, x^a) \right)+ \frac{6\pi \ell^4}{c r^2}\   \delta T_{ab}(r, x^a)  = {\cal O}\big((\ell/r)^4\big)\, .
\end{equation} 
This remarkable relation demonstrates that the boundary condition dynamically rotates along the RG flow. What begins as a purely Dirichlet condition at the asymptotic boundary  evolves into a mixed (Robin-like) Neumann-Dirichlet condition on the finite-cutoff metric $\gamma_{ab}$ \cite{Compere:2008us, Guica:2019nzm, Adami:2025pqr, Sheikh-Jabbari:2026uol, Parvizi:2025wsg}. 

Two features of this mixed condition stand out. First, the Dirichlet part is modified by the Weyl factor $e^{(\Phi-1)/2}$. The presence of this factor is a direct consequence of leaving the lapse function $\Phi$ arbitrary, differentiating our setup from standard Fefferman-Graham analyses. Second, the admixture with the Neumann condition ($\delta T_{ab}$) is explicitly suppressed by the central charge $1/c$ and by the radial running factor $\ell^2/r^2$. Therefore, as we evolve more into the IR (smaller $r$ values) the Neumann admixture increases and we distance further from pure Dirichlet at large $r$.

\subsection{Physical implications of the radial flow of boundary conditions}\label{sec:recap-bc-evol}
We now state the central argument regarding the radial evolution of the boundary action. The algebraic mixing of metric and stress-tensor fluctuations observed in the previous subsection has a direct and elegant interpretation at the level of the variational principle.

\paragraph{Variational principle along the flow.}
As established by the deformation flow equation \eqref{d-f-NR-D-Gr}, the effective action of the boundary theory at a finite cutoff $\Sigma_r$ is related to the action at the asymptotic boundary $\Sigma$ by integrating the RG flow. Formally, this integration yields an accumulated deformation term \cite{Parvizi:2025wsg, Sheikh-Jabbari:2025kjd}:   
\begin{equation}
    {S}^{\text{\tiny{D}}}_{\text{bdry}}[\Sigma_r]  =  {S}^{\text{\tiny{D}}}_{\text{bdry}}[\Sigma]  + {S}_{\text{\tiny{deform}}}\, ,
\end{equation}
where ${S}_{\text{\tiny{deform}}}$ represents the exact integral of the flow equation from infinity down to the radial slice $r$. While we will explicitly compute ${S}_{\text{\tiny{deform}}}$ in section \ref{sec:Int-bdry-action} and reveal that it contains the induced JT gravity action, its mere formal presence is sufficient to dictate the evolution of the boundary conditions. To understand these physical implications, consider the general variation of this relationship:
\begin{equation}
    \delta{S}^{\text{\tiny{D}}}_{\text{bdry}}[\Sigma_r]  =  \delta{S}^{\text{\tiny{D}}}_{\text{bdry}}[\Sigma]  + \delta {S}_{\text{\tiny{deform}}}\, .
\end{equation}
This simple equation governs how boundary conditions are dynamically exchanged between the UV (asymptotic) and the IR (finite cutoff) regimes.

\paragraph{Flowing from the UV to the IR.}
If we impose a strict Dirichlet boundary condition at the asymptotic boundary $\Sigma$, we demand, by definition, that $\delta{S}^{\text{\tiny{D}}}_{\text{bdry}}[\Sigma] = 0$. Consequently, evaluating the variation at the finite cutoff surface $\Sigma_r$ yields:
\begin{equation}
   0 \neq  \delta{S}^{\text{\tiny{D}}}_{\text{bdry}}[\Sigma_r]  =  \delta {S}_{\text{\tiny{deform}}}\, .
\end{equation}
This explicitly demonstrates that radial evolution, as governed by the bulk equations of motion, strictly breaks the Dirichlet boundary condition at finite distances. Instead, the well-posed variational principle at $\Sigma_r$ dictates:
\begin{equation}\label{grav-mat-Sigma-r}
    \delta \left( {S}^{\text{\tiny{D}}}_{\text{bdry}}[\Sigma_r]  -  {S}_{\text{\tiny{deform}}} \right) = 0\, . 
\end{equation}
This relation implies a dynamically mixed boundary condition on $\Sigma_r$. Both the induced metric and its conjugate stress tensor are free to fluctuate, but their variations are not independent—they are rigidly constrained by $\delta S_{\text{\tiny{deform}}}$. Physically, this describes a constrained ``gravity coupled to matter'' system residing at a finite radial distance, governed by the effective action ${S}^{\text{\tiny{D}}}_{\text{bdry}}[\Sigma_r]$.

\paragraph{Flowing from the IR to the UV.}
Conversely, one might attempt to work in the opposite direction by forcefully imposing a strict Dirichlet boundary condition at a finite distance, thereby requiring $\delta {S}^{\text{\tiny{D}}}_{\text{bdry}}[\Sigma_r] = 0$. In this scenario, the boundary condition at asymptotic infinity $\Sigma$ can no longer be Dirichlet:
\begin{equation}
      0 \neq \delta{S}^{\text{\tiny{D}}}_{\text{bdry}}[\Sigma]  = - \delta {S}_{\text{\tiny{deform}}}\, ,
\end{equation}
which is equivalent to demanding:
\begin{equation}\label{grav-mat-Sigma}
   \delta \left( {S}^{\text{\tiny{D}}}_{\text{bdry}}[\Sigma]  + {S}_{\text{\tiny{deform}}} \right) = 0\, .
\end{equation}
This generically leads to a mixed boundary condition at the asymptotic boundary $\Sigma$, describing an alternative effective gravity-matter system at infinity. 

We close this part by the remark that the above discussions on the flow of boundary conditions and the analysis in section \ref{sec:derivation-bulk-eom} about the consistency of the analysis, are complementary to each other: the flow of boundary conditions is a result of the bulk equations of motion.

\paragraph{The physical origin of finite-cutoff instabilities.}
A crucial observation emerges from comparing the effective variational principles in \eqref{grav-mat-Sigma-r} and \eqref{grav-mat-Sigma}: there is a relative sign difference in front of the deformation term $S_{\text{\tiny{deform}}}$. 

The term $S_{\text{\tiny{deform}}}$ is directly responsible for activating the gravitational dynamics at the boundary. Thus, this sign reversal dictates that the induced gravitational action will carry opposite overall signs in the two scenarios. Consequently, only one of these setups can yield a well-defined effective theory with healthy (positive-definite) kinetic terms for the boundary metric. This sign discrepancy provides a profound physical origin for the generic ill-posedness \cite{Avramidi:1997sh, Avramidi:1997hy, anderson2008boundary, Marolf:2012dr, Figueras:2011va, Witten:2018lgb, Liu:2024ymn, Banihashemi:2024yye}—such as the inevitable appearance of ghost-like instabilities—that plagues attempts to arbitrarily impose strict Dirichlet boundary conditions at a finite radial cutoff.

\section{Emergence of 2d gravity: Integrating the flow equation}\label{sec:Int-bdry-action}
Having established the differential RG flow equation and its profound implications for boundary conditions, we now turn to our primary objective: integrating this flow equation to uncover the exact effective action of the finite-cutoff theory. In this section, we show how the radial evolution inevitably induces a 2d scalar-tensor gravity theory—specifically, Jackiw-Teitelboim (JT) gravity—on the finite cutoff slice.

\subsection{Recalling the statement of AdS/CFT}\label{sec:AdS-CFT-recap}

Before constructing an exact, local solution to the deformation flow equation \eqref{d-f-NR-D-Gr}, it is highly instructive to first recall the standard AdS/CFT correspondence in the saddle-point approximation. This will reveal the presence of JT gravity in the asymptotic expansion.

\paragraph{The holographic dictionary.}
According to the standard holographic dictionary \cite{Gubser:1998bc, Witten:1998qj} (see \cite{Parvizi:2025shq, Parvizi:2025wsg} and references therein), the effective action of the dual quantum field theory at a given cutoff surface is equivalent to the on-shell gravitational action evaluated in the enclosed bulk region. For a finite cutoff surface $\Sigma_r$ and the true asymptotic boundary $\Sigma$, this implies:
\begin{equation}\label{AdS/CFT-r}
    \begin{split}
        & {S}^{\text{\tiny{D}}}_{\text{bdry}}[\Sigma_r] = S^{\text{\tiny{D}}}_{\text{\tiny{bulk}}}[\mathcal{M}_r]\Big|_{\text{\tiny{on-shell}}} \, , \\
        & {S}^{\text{\tiny{D}}}_{\text{bdry}}[\Sigma] = S^{\text{\tiny{D}}}_{\text{\tiny{bulk}}}[\mathcal{M}]\Big|_{\text{\tiny{on-shell}}}\, ,
    \end{split}
\end{equation}
where the superscript D indicates that Dirichlet boundary conditions are imposed on the respective surfaces. 

\paragraph{Localizing the bulk action.}
To evaluate these expressions, we substitute the bulk equations of motion into the action \eqref{action-AdS}. Using the constant scalar curvature $\mathscr{R}=-6/\ell^2$, the volume element relation $\sqrt{-g}=\frac{r}{\ell}\Phi \sqrt{-\gamma}$, and the definition of the stress tensor trace \eqref{BY-EMT-Kab}, the on-shell bulk action reduces to:
\begin{equation}
    S^{\text{\tiny{D}}}_{\text{\tiny{bulk}}}[\mathcal{M}_r]\Big|_{\text{\tiny{on-shell}}} = -\frac{2r}{\kappa_3 \ell^3} \int_{\mathcal{M}_r} \d{}^{3}x\, \sqrt{-\gamma}\, \Phi -  \int_{\Sigma_r}\d{}^{2}x\, \sqrt{-\gamma}\, T \, .
\end{equation} 
To write the remaining bulk integral strictly as a boundary term, we can employ a mathematical trick by introducing an auxiliary scalar field $\chi$ defined via the differential equation:
\begin{equation}\label{chi-eq}
    \mathcal{D}_r (\sqrt{-\gamma}\, \chi) =  \frac{\sqrt{-\gamma}\, \Phi}{\kappa_3 }\frac{r}{\ell}\, . 
\end{equation}
By replacing the bulk integrand with this total radial derivative, the bulk on-shell action reduces entirely to a boundary integral on $\Sigma_r$:
\begin{equation}\label{Sbdry-r}
    S^{\text{\tiny{D}}}_{\text{\tiny{bulk}}}[\mathcal{M}_r]\Big|_{\text{\tiny{on-shell}}} = - \int_{\Sigma_r} \d{}^{2}x\, \sqrt{-\gamma} \left( T + \frac{2}{\ell^2}\chi \right)\, .
\end{equation}

\paragraph{Relating the UV and IR actions.}
We now seek the relationship between the field theory actions at the two respective boundaries, ${S}^{\text{\tiny{D}}}_{\text{bdry}}[\Sigma_r]$ and ${S}^{\text{\tiny{D}}}_{\text{bdry}}[\Sigma]$. Subtracting the two relations in \eqref{AdS/CFT-r} and utilizing \eqref{Sbdry-r}, we obtain:
\begin{equation}\label{Sbdry-soln}
    {S}^{\text{\tiny{D}}}_{\text{bdry}}[\Sigma_r] = {S}^{\text{\tiny{D}}}_{\text{bdry}}[\Sigma] - \int_{\Sigma_r} \d{}^{2}x\, \sqrt{-\gamma} \left( T + \frac{2}{\ell^2}\chi \right) + \int_{\Sigma} \d{}^{2}x\, \sqrt{-\gamma} \left( T + \frac{2}{\ell^2}\chi \right)\, .
\end{equation}
One can readily verify that \eqref{Sbdry-soln} is a valid mathematical solution to the differential flow equation \eqref{d-f-NR-D-Gr}, provided the scalar field $\chi$ satisfies \eqref{chi-eq}. 

\paragraph{Asymptotic expansion and induced JT gravity.}
While \eqref{Sbdry-soln} is exact, its right-hand side explicitly mixes quantities evaluated at two different surfaces ($\Sigma$ and $\Sigma_r$). For a coherent physical interpretation, we must express the action exclusively in terms of variables defined at a single surface. Using the asymptotic expansions derived in section \ref{sec:asymptotic-solution-space}, we can evaluate the $\Sigma$ integral. From the asymptotic properties \eqref{asymp-const}, we have:
\begin{equation}
 \sqrt{-\gamma}\Big|_{\Sigma} = \sqrt{-q}\, , \qquad   T\Big|_{\Sigma} = -\frac{c}{6\pi \ell^2} \frac{r_\infty^2}{\ell^2} -\frac{c R[q]}{24\pi}\, , \qquad \chi\Big|_{\Sigma} = \frac{c r_\infty^2}{24\pi \ell^2} - \frac{c \ell^2}{48\pi}R[q] \ln(r/\ell)\, .
\end{equation}
Substituting these along with the finite-$r$ expansions \eqref{asymp-grav-quantities} into \eqref{Sbdry-soln}, dropping purely topological and total derivative terms, we are left with the following simplified expression:
\begin{equation}\label{Sbdry-Sigma_r-Sigma}
    \begin{split}
      \hspace{-4mm}   {S}^{\text{\tiny{D}}}_{\text{bdry}}[\Sigma_r] = {S}^{\text{\tiny{D}}}_{\text{bdry}}[\Sigma]  + \int_{\Sigma} \d{}^{2}x\, \sqrt{-q}  \Bigg[  \frac{c(r^2 - r_\infty^2)}{12\pi \ell^4} + \frac{\ell^2}{r^2}\bigg( \frac{3\pi \ell^2}{2c}\ttbarb + \frac{c(\Phi_1^2 - \Phi_2)}{24\pi \ell^2} - \frac{c \Phi_1}{48\pi} R[q] \bigg)\Bigg] + \mathcal{O}\Big(\frac{\ell^4}{r^4}\Big).
    \end{split}
\end{equation}

This equation allows us to clearly interpret the physical mechanics of the flow. Suppose we impose the strict Dirichlet boundary condition $\delta \gamma_{ab}|_{\Sigma_r}=0$ at the finite cutoff $\Sigma_r$, which guarantees a pure, non-gravitational QFT resides there. Eq.  \eqref{Sbdry-Sigma_r-Sigma} dictates that doing so forces us to relax Dirichlet conditions at asymptotic infinity $\Sigma$. Consequently, the UV theory at infinity is no longer a pure QFT; metric fluctuations $\delta q_{ab} \neq 0$ are activated. 

Crucially, the deformation terms in \eqref{Sbdry-Sigma_r-Sigma} explicitly show that the UV theory is modified not only by matter-sector deformations (the standard $\ttbarb$ term) but also by gravitational contributions. The most notable term is the $\Phi_1 R[q]$ coupling, which famously corresponds to Jackiw-Teitelboim (JT) gravity \cite{Jackiw:1984je, Teitelboim:1983ux} living at the asymptotic boundary. 

\paragraph{The Fefferman-Graham limit.}
It is worth noting what happens if we adopt the standard Fefferman-Graham gauge by fixing $\Phi_i=0$. Under this restriction, the gravitational dilaton terms vanish, and the boundary action \eqref{Sbdry-Sigma_r-Sigma} dramatically simplifies to:
\begin{equation}\label{Sbdry-Sigma_r-Sigma-FG}
    \begin{split}
      {S}^{\text{\tiny{D}}}_{\text{bdry}}[\Sigma_r] = {S}^{\text{\tiny{D}}}_{\text{bdry}}[\Sigma]  + \int_{\Sigma} \d{}^{2}x\, \sqrt{-q}  \Bigg[  \frac{c(r^2 - r_\infty^2)}{12\pi \ell^4} + \frac{3\pi \ell^4}{2c r^2}\ttbarb  \Bigg]\, .
    \end{split}
\end{equation}
In this gauge, the JT term drops out, leaving only the pure matter T$\bar{\text{T}}$ modification \cite{Smirnov:2016lqw, Cavaglia:2016oda, McGough:2016lol}. Note that while \eqref{Sbdry-Sigma_r-Sigma} is a perturbative result in the asymptotic expansion, the FG result \eqref{Sbdry-Sigma_r-Sigma-FG} is exact.

\paragraph{Motivation for a local ansatz.}
While the preceding analysis elegantly illustrates how imposing a Dirichlet condition at a finite cutoff induces JT gravity in the UV, our primary goal is the exact reverse: we want to impose strict Dirichlet boundary conditions at asymptotic infinity and investigate the induced gravity on the boundary QFT at a finite distance. To achieve this cleanly, we need an exact, non-perturbative solution to the flow equation formulated entirely in terms of local geometric variables strictly residing on the cutoff surface $\Sigma_r$. We develop this targeted formulation in the subsequent subsection.

\subsection{Induced JT gravity at finite cut-off}\label{sec:finite-cutoff-action}
The asymptotic  analysis in the previous subsection revealed the presence of JT gravity near the boundary $\Sigma$. We now seek an exact, non-perturbative solution to the deformation flow equation \eqref{d-f-NR-D-Gr}. Our goal is to construct a boundary action formulated entirely in terms of local quantities residing on the cutoff surface $\Sigma_r$.

\paragraph{The JT gravity ansatz.}
Motivated by the structure of the flow equation \eqref{d-f-NR-D-Gr} and the form of the asymptotic result \eqref{Sbdry-Sigma_r-Sigma}, we propose the following ansatz for the finite-cutoff boundary action:
\begin{equation}\label{Asnzatz-T-log-renormalized}
\inbox{{S}^{\text{\tiny{D}}}_{\text{bdry}}[\Sigma_r]
= {S}^{\text{\tiny{D}}}_{\text{bdry}}[\Sigma]
- \frac{1}{2} \int_{\Sigma_r}\d{}^{2}x\, \sqrt{-\gamma}\,  T
+ \frac{c}{12\pi } \int_{\Sigma_r} \d{}^{2}x\, \sqrt{-\gamma}\,  
\Psi\, \big(R[\gamma] - 2 \Lambda(r) \big)\, .}
\end{equation}
In this expression, $T$ is the trace of the stress tensor, $\Psi(r, x^a)$ is a dynamical scalar field playing the role of the JT dilaton, and $\Lambda(r)$ is an effective 2d cosmological constant. The second term on the right-hand side, as we will discuss in section \ref{sec:Neumann-interp},  acts as a generator for a Legendre transformation \cite{Compere:2008us, Parvizi:2025shq}.

\paragraph{Localization of the flow equation.}
To verify if this ansatz satisfies the flow equation \eqref{d-f-NR-D-Gr}, we substitute \eqref{Asnzatz-T-log-renormalized} into the radial derivative. After discarding total derivative terms on the spatial slice $\Sigma_r$, we obtain the following integral constraint:
\begin{equation}
\int_{\Sigma_r} \d{}^{2}x\, \sqrt{-\gamma} 
\Bigg\{
\frac{c}{12 \pi } (R[\gamma] - 2\Lambda) \partial_{r}\Psi
+ \frac{c \Psi}{6\pi r}(2\Lambda - r\Lambda')
- \frac{c \Phi}{24\pi r}R[\gamma]
+ \frac{2\ell^4 }{r^3}\Psi 
\big(T^{ab}\nabla_{a}\Phi \nabla_{b}\Phi + \Lambda \Phi T\big)
\Bigg\} = 0 .
\end{equation}
To extract a local physical theory, we demand that the integrand vanishes up to a total divergence. A natural choice is to set the integrand equal to $-\frac{c}{24\pi r} R[\gamma]$, which yields the local dilaton evolution equation:
\begin{equation}\label{Basic-Psi-eq}
\frac{c}{12 \pi } (R[\gamma] - 2\Lambda) \partial_{r}\Psi
+ \frac{c \Psi}{6\pi r}(2\Lambda - r\Lambda')
- \frac{c \Phi}{24\pi r}R[\gamma]
+ \frac{2\ell^4 }{r^3}\Psi 
\big(T^{ab}\nabla_{a}\Phi \nabla_{b}\Phi + \Lambda \Phi T\big)
= - \frac{c}{24\pi r} R[\gamma] .
\end{equation}

\paragraph{Dilaton RG flow equation.}
Eq. \eqref{Basic-Psi-eq} can be recast as a first-order linear differential equation for the dilaton field $\Psi$:
\begin{equation}\label{Psi-eq}
\partial_r\Psi + \mathcal{X}\Psi = \mathcal{S},
\end{equation}
where the ``drag'' coefficient $\mathcal{X}$ and the ``source'' term $\mathcal{S}$ are given by:
\begin{equation}\label{X-S-non-D}
\begin{split}
\mathcal{X} &=
\frac{
\frac{4\Lambda}{r}\!\left( 1 + \frac{6\pi \ell^4}{c r^2} \Phi\, T \right)
-2 \Lambda'
+ \frac{24\pi \ell^4}{c r^3} T^{ab}\nabla_{a}\nabla_{b}\Phi
}{R[\gamma]-2\Lambda},
\\
\mathcal{S} &=
\frac{\Phi-1}{2r}\,
\frac{ R[\gamma] }{R[\gamma]-2\Lambda}.
\end{split}
\end{equation}
The source term $\mathcal{S}$ is notably proportional to $(\Phi-1)$, confirming that the non-trivial radial lapse $\Phi$ is the primary driver of the dilaton's radial evolution. Using an integrating factor $\mu(r,x^a)$, the exact solution for the dilaton is:
\begin{equation}\label{Psi-soln}
\Psi(r,x^a) =
\frac{1}{\mu(r,x^a)}\left(
\int^{r} \d{r}'\, \mu(r',x^a)\, \mathcal{S}(r',x^a)
+ {\Psi_0(x^a)}\right) ,
\quad
\mu(r,x^a)=
\exp\!\left(
\int^{r} \d r' \, \mathcal{X}(r',x^a)
\right).
\end{equation}
Here, $\Psi_0(x^a)$ is an integration constant independent of the radial coordinate.

\paragraph{Radial dependence and consistent expansions.}
The function $\Lambda(r)$ remains a free parameter of the integration. To match the expected UV behavior where gravity becomes subdominant, a careful analysis of the solution \eqref{Psi-soln} reveals that we must assume the following expansions for the cosmological constant and the dilaton field:
\begin{equation}\label{Psi-fall-off}
\begin{split}
\Lambda(r)
&=
 \left(\frac{r}{\ell}\right)^2\ \Lambda_0
+ \Lambda_1
+  \left(\frac{\ell}{r}\right)^2\ \Lambda_2
+ \cdots ,\qquad \Lambda_0\neq 0\, , \\
\Psi(r,x^a)
&=
\left(\frac{\ell}{r}\right)^2\ \Psi_0(x^a)
+ \left(\frac{\ell}{r}\right)^4\ \Psi_1(x^a)
+ \left(\frac{\ell}{r}\right)^6\ \Psi_2(x^a)
+ \cdots \, .
\end{split}
\end{equation}

\paragraph{Iterative solution for dilaton coefficients.}
By substituting these expansions into the ODE \eqref{Psi-eq} and using the semi-on-shell expansions of the gravitational variables from \eqref{asymp-grav-quantities}, we solve for the dilaton coefficients iteratively. The leading-order coefficient $\Psi_0$ is fixed by the requirement that the boundary action successfully reduces to the undeformed action as $r \to \infty$:
\begin{equation}\label{Psi0-Phi1}
\inbox{\Psi_0 = - \frac{\Phi_1}{2\ell^2 \Lambda_0}.} 
\end{equation}
Continuing the iteration, the first two subleading coefficients are determined as:
\begin{equation}
\begin{split}
\Psi_1 &= - \frac{\Phi_1}{4\ell^2 \Lambda_0^2} \left(R[q] -2\Lambda_1 + 2\Lambda_0 \Phi_1\right), \\ 
\Psi_2  &= - \frac{1}{\Lambda_0} \Bigg[ \frac{\Phi_1^3}{4\ell^2} + \frac{\Phi_1 \Phi_2}{4 \ell^2} - \frac{\Phi_2 }{16} R[q] + \frac{\Phi_1}{16} D^2 \Phi_1 + \frac{9 \pi^2 \ell^2}{c^2} \Phi_1 \ttbarb - \frac{\Lambda_0 \Phi_1^2}{ 2\ell^2 \Lambda_0} + \frac{\Lambda_1^2 + \Lambda_2 \Lambda_0}{2\ell^2 \Lambda_0^2} \Phi_1 \\
& \quad + \frac{3}{16} \left(1 + \frac{2}{\ell^2 \Lambda_2}\right) \Phi_1^2 R[q] - \frac{\Lambda_1 \Phi_1 R[q]}{2\ell^2 \Lambda_0^2} + \frac{8 + \ell^2 \Lambda_0 (2 + \ell^2 \Lambda_0)}{64\ell^2 \Lambda_0^2} \Phi_1 R[q]^2 \Bigg].
\end{split}
\end{equation}

\paragraph{Verification via the boundary action expansion.}
With the iterative solutions in hand, we evaluate the asymptotic expansion of the full boundary action \eqref{Asnzatz-T-log-renormalized}. To the first non-trivial order, the action becomes:
\begin{equation}\label{Sbdry-expansion-check}
    \begin{split}
     {S}^{\text{\tiny{D}}}_{\text{bdry}}[\Sigma_r] = {S}^{\text{\tiny{D}}}_{\text{bdry}}[\Sigma]  + \int_{\Sigma} \d{}^{2}x\, \sqrt{-q}  \Bigg[  \frac{c r^2 }{12\pi \ell^4} + \frac{\ell^2}{r^2}\Bigg( \frac{3\pi \ell^2}{2c}\ttbarb + \frac{c(\Phi_1^2 - \Phi_2)}{24\pi \ell^2} - \frac{c \Phi_1}{48\pi} R[q] \Bigg)\Bigg] + \mathcal{O}\Big(\frac{\ell^4}{r^4}\Big).
    \end{split}
\end{equation}
This expansion perfectly matches the result obtained from the direct holographic dictionary in \eqref{Sbdry-Sigma_r-Sigma}. This agreement provides robust evidence that our local JT ansatz is correct, confirming that the bulk radial lapse function $\Phi$ is the unique physical source for the induced JT gravity on the finite-cutoff surface.

\subsection{Induced JT gravity with Neumann boundary conditions}\label{sec:Neumann-interp}
The presence of the term $- \frac{1}{2} \int_{\Sigma_r}\d{}^{2}x\, \sqrt{-\gamma}\, T$ in our ansatz \eqref{Asnzatz-T-log-renormalized} warrants a deeper physical clarification. As we will demonstrate, this term is not merely a technical addition but serves as the generator of a canonical transformation between boundary conditions \cite{Papadimitriou:2010as, Papadimitriou:2016yit}.

\paragraph{Legendre transformation and the Neumann action.}
In the context of the holographic variational principle, this term allows  transition from the Dirichlet boundary action, ${S}^{\text{\tiny{D}}}_{\text{bdry}}[\Sigma_r]$, to a new boundary action defined as \cite{Parvizi:2025shq, Parvizi:2025wsg, Sheikh-Jabbari:2025kjd}:
\begin{equation}\label{S-N-def}
    {S}^{\text{\tiny{N}}}_{\text{bdry}}[\Sigma_r] = {S}^{\text{\tiny{D}}}_{\text{bdry}}[\Sigma_r] + \frac{1}{2}\int_{\Sigma_r}\d{}^{2}x\, \sqrt{-\gamma}\, T\, .
\end{equation}
The physical significance of ${S}^{\text{\tiny{N}}}_{\text{bdry}}$ becomes apparent when we evaluate its variation. Recalling that the variation of ${S}^{\text{\tiny{D}}}_{\text{bdry}}$ is driven by the stress tensor \eqref{dr-delta-S-2}, the variation of the new action \eqref{S-N-def} takes the form:
\begin{equation}
    \delta {S}^{\text{\tiny{N}}}_{\text{bdry}}[\Sigma_r] = \frac{1}{2} \int_{\Sigma_r}\d{}^{2}x\, \delta(\sqrt{-\gamma}\, T^{ab}) \, \gamma_{ab}\, .
\end{equation}
This variation confirms that the action ${S}^{\text{\tiny{N}}}_{\text{bdry}}$ is compatible with Neumann boundary conditions, defined by $\delta(\sqrt{-\gamma}\, T^{ab}) = 0$. Under this condition, the variation of the action vanishes identically, rendering the variational principle well-posed for a fixed stress tensor \cite{Compere:2008us, Krishnan:2016mcj}.

\paragraph{The Neumann-JT formulation.}
Building on this insight, we can suggestively rewrite our exact solution \eqref{Asnzatz-T-log-renormalized} by absorbing the Legendre term into the left-hand side. This allows us to express the result as an effective Neumann action:
\begin{equation}\label{N-bdry-Sigma-r-unreorm}
\inbox{{S}^{\text{\tiny{N}}}_{\text{bdry}}[\Sigma_r]
= {S}^{\text{\tiny{D}}}_{\text{bdry}}[\Sigma]
+ \frac{c}{12\pi } \int_{\Sigma_r} \d{}^{2}x\, \sqrt{-\gamma}\,  
\Psi\, \big(R[\gamma] - 2 \Lambda(r) \big)\, .}
\end{equation}

\paragraph{Physical interpretation.}
This formulation offers a compelling physical narrative for the GR from RG proposal: Holographic RG induces a Neumann boundary condition at a finite distance originates directly from a Dirichlet boundary condition at the AdS conformal boundary. In this picture, the Dirichlet action at infinity acts as the ``initial condition'' for the flow. As the energy scale is lowered (moving into the bulk), the integrated JT action acts as a dynamical operator that ``rotates'' the boundary condition, naturally yielding a Neumann condition at the finite-radius surface $\Sigma_r$. Note that the Neumann boundary condition means that variations of 2d metric $\delta\gamma_{ab}$ is not restricted by any boundary condition (induced from the 3d bulk description), and that having an unrestricted $\delta\gamma_{ab}$ is what is required by the standard 2d gravity description. 


\section{Renormalized holographic RG flow}\label{sec:renormalization}
In the previous sections, we analyzed the radial flow of the unrenormalized (bare) gravitational action. While this provided a clear local derivation of the induced JT dynamics, a complete holographic description requires handling the infinite volume divergences of AdS spacetime. In this section, we extend our framework to the renormalized holographic action. We demonstrate that the emergence of induced 2d gravity is a robust feature of the theory that survives the standard holographic renormalization procedure \cite{deHaro:2000vlm, Skenderis:2002wp, Bianchi:2001kw}, resulting in a finite and physically well-defined boundary effective action.

\subsection{Renormalized Dirichlet deformation flow equation}\label{sec:deformation-flow-D-R}

\paragraph{Renormalized action and stress tensor.}
Boundary observables are rendered finite by supplementing the Einstein-Hilbert action with the standard holographic counterterm proportional to the boundary volume element \cite{Balasubramanian:1999re, Emparan:1999pm}. The renormalized action for a bulk region $\mathcal{M}_r$ with Dirichlet boundary conditions is:
\begin{equation}\label{action-AdS-R-D}
    \mathcal{S}^{\text{\tiny{D}}}_{\text{\tiny{bulk}}}[\mathcal{M}_r] = \frac{1}{2\kappa_3} \int_{\mathcal{M}_r} \d{}^{3}x\, \sqrt{-g} \left(\mathscr{R} + \frac{2}{\ell^2}\right) + \frac{r^2}{\ell^2 \kappa_3}\int_{\Sigma_r} \d{}^{2}x\, \sqrt{-\gamma}\, \left(K - \frac{1}{\ell}\right)\, .
\end{equation} 
This inclusion does not alter the Dirichlet nature of the variational principle; rather, it shifts the momentum conjugate to the boundary metric. The resulting renormalized Brown-York energy-momentum tensor (rBY-EMT)—often called the holographic stress tensor—is given by \cite{Balasubramanian:1999re, deHaro:2000vlm, Skenderis:2002wp}:
\begin{equation}\label{r-BY-EMT}
    \mathcal{T}_{ab}=\frac{1}{\kappa_3}\left(K_{ab} - \frac{r^2}{\ell^2} K \gamma_{ab} + \frac{r^2}{\ell^3} \gamma_{ab}\right) = T_{ab} + \frac{c}{12\pi}\frac{r^2}{\ell^4} \gamma_{ab}\, .
\end{equation}  
While the bare tensor $T_{ab}$ diverges as $r \to \infty$, the holographic tensor $\mathcal{T}_{ab}$ remains finite at the boundary. Specifically, its asymptotic limit is related to the dual QFT stress tensor as $\mathcal{T}_{ab} \to - t_{ab}$ (cf. \eqref{asymp-const}, \eqref{asymp-grav-quantities}), which is compatible with the results in the FG gauge \cite{Balasubramanian:1999re, deHaro:2000vlm}.

\paragraph{Renormalized bulk dynamics.}
Projecting the Einstein equations into the radial foliation using the renormalized variables \eqref{r-BY-EMT} yields the following set of constrained evolution equations: 
\begin{subequations}\label{EoM-GR-mat-decompose}
    \begin{align}
       & \mathcal{T} + \frac{{c} }{24\pi} R + \frac{6\pi\ell^4}{{c r^2}}\Ottbar  = 0 \, ,\label{EoM-ss-ren}
       \\
       & {\nabla}_b{\mathcal{T}}^{b}_{a}=0\, , \label{EoM-sa-ren}\\
       & \Phi^{-1}\mathcal{D}_r{\mathcal{T}}_{ab}  - \frac{12 \pi \ell^4}{c r^3} \left(\mathcal{T} \mathcal{T}_{ab} + \frac{3}{2} \Ottbar \gamma_{ab}\right) + \frac{c}{12 \pi r} \left( \frac{\nabla_{a} \nabla_{b}\Phi}{\Phi} - \frac{\Box \Phi}{\Phi} \gamma_{ab} \right)=0\, . \label{EoM-ab-ren}
    \end{align}
\end{subequations}  
where $\Ottbar := \mathcal{T}^{ab}\mathcal{T}_{ab} - \mathcal{T}^2$. These equations demonstrate that the quadratic T$\bar{\text{T}}$ structure persists in the renormalized theory, though with radial weights that differ from the bare case.

\paragraph{Derivation of the renormalized flow.}
Following the same logic as in section \ref{sec:deformation-flow-D}, we evaluate the radial derivative of the renormalized on-shell action by considering its variation under the radial diffeomorphism $\xi = \partial_r$:
\begin{equation}
   \begin{split}
       \frac{\d{}}{\d{}r} \mathcal{S}^{\text{\tiny{D}}}_{\text{bulk}}[\mathcal{M}_r] \Big|_{\text{on-shell}} & = -\frac{\ell^2}{2r^2} \int_{\Sigma_r} \sqrt{-\gamma}\, \mathcal{T}^{ab}\, \partial_r \left( \frac{r^2}{\ell^2}\gamma_{ab} \right)\\
       & = - \int_{\Sigma_r} \sqrt{-\gamma}\, \Phi\, \left( \frac{12 \pi \ell^4}{c r^3} \Ottbar + \frac{\mathcal{T}}{r} \right)\, .
   \end{split}
\end{equation}
Invoking the holographic dictionary, we obtain the renormalized deformation flow equation \cite{Hartman:2018tkw, Parvizi:2025wsg, Sheikh-Jabbari:2025kjd}:
\begin{equation}\label{TTbar-deformation-flow}
    \inbox{\frac{\d{}}{\d{}r} \mathcal{S}^{\text{\tiny{D}}}_{\text{bdry}}[\Sigma_r] = - \int_{\Sigma_r} \sqrt{-\gamma}\, \Phi\, \left( \frac{12 \pi \ell^4}{c r^3} \Ottbar + \frac{\mathcal{T}}{r} \right)\, .}
\end{equation}
Using the renormalized Hamiltonian constraint \eqref{EoM-ss-ren} to replace the $\Ottbar$ term with the boundary Ricci scalar, we find the equivalent geometric representation of the renormalized flow:
\begin{equation}\label{Ren-D-flow-eq}
   \inbox{ \frac{\d{}}{\d{}r} \mathcal{S}^{\text{\tiny{D}}}_{\text{bdry}}[\Sigma_r] = \frac{c}{12\pi r}\int_{\Sigma_r} \sqrt{-\gamma}\, \Phi\, \left( R + \frac{12\pi}{c}\mathcal{T}\right)\, .}
\end{equation}
Eq. \eqref{Ren-D-flow-eq} reveals that even after renormalization, the radial RG flow maintains a JT-like structure, with the lapse $\Phi$ sourcing a curvature term $R$, now accompanied by a trace contribution from the holographic stress tensor.

\subsection{Integrating the renormalized action flow equation}\label{subsec:int-renorm-bdry}
With the renormalized flow equation \eqref{Ren-D-flow-eq} established, we now integrate it to find the exact local boundary action. As in the bare case, we seek a solution where the non-trivial radial lapse $\Phi$ seeds a gravitational dilaton field.

\paragraph{Renormalized JT ansatz.}
We consider the following ansatz for the renormalized boundary action at the cutoff surface $\Sigma_r$:
\begin{equation}\label{S-bdry-alpha-beta}
\inbox{\mathcal{S}^{\text{\tiny{D}}}_{\text{bdry}}[\Sigma_r]
= \mathcal{S}^{\text{\tiny{D}}}_{\text{bdry}}[\Sigma]
+ \frac{c}{12\pi } \int_{\Sigma_r} \d{}^{2}x\, \sqrt{-\gamma}\,  
\Psi\, \big(R[\gamma] - 2 \Lambda(r) \big)\, .}
\end{equation}
Note that, unlike the unrenormalized ansatz \eqref{Asnzatz-T-log-renormalized}, we do not include an explicit $T$ term here; the counterterms already present in the renormalized bulk action \eqref{action-AdS-R-D} naturally account for the trace structure.

\paragraph{Dilaton RG flow equation.}
Substituting this ansatz into the renormalized flow equation \eqref{Ren-D-flow-eq} yields an integral constraint. Following the localization procedure—equating the integrand to the purely topological term $-\frac{c}{24\pi r}R[\gamma]$—we obtain a first-order linear differential equation for the renormalized dilaton $\Psi$:
\begin{equation}\label{Psi-eq-2}
    \partial_{r}\Psi + \mathcal{X} \Psi = \mathcal{S}\, .
\end{equation}
For the renormalized flow, the ``drag'' and ``source'' coefficients are given by:
\begin{align}\label{X-S-renorm}
    \mathcal{X} &= \frac{1}{r} \frac{ 4\Lambda\left[ 1+ \left(\frac{6\pi \ell^4}{c r^2} \mathcal{T} - 1\right)\Phi \right] + \frac{24\pi \ell^4}{c r^2} \mathcal{T}^{ab} \nabla_{a}\nabla_{b}\Phi - 2 (\Box \Phi + r \Lambda') }{R[\gamma] - 2 \Lambda}\, , \\
    \mathcal{S} &= \frac{1}{2r} \frac{(2\Phi-1)R[\gamma] + \frac{24\pi}{c} \Phi \mathcal{T} }{R[\gamma]-2\Lambda}\, .
\end{align}
The source term $\mathcal{S}$ is now driven by both the curvature $R$ and the holographic trace $\mathcal{T}$, weighted by the lapse function.

\paragraph{Asymptotic matching and the UV dilaton.}
We solve \eqref{Psi-eq-2} using the same asymptotic expansions for $\Psi$ and $\Lambda$ as in \eqref{Psi-fall-off}. Near the boundary, the induced action behaves as:
\begin{equation}
    \mathcal{S}^{\text{\tiny{D}}}_{\text{bdry}}[\Sigma_r]
= \mathcal{S}^{\text{\tiny{D}}}_{\text{bdry}}[\Sigma]
- \frac{c \Lambda_0}{6\pi } \int_{\Sigma_r} \d{}^{2}x\, \sqrt{-q}\, \Psi_0\, .
\end{equation}
To ensure that the cutoff action successfully reduces to the undeformed boundary action as $r \to \infty$, we must set the integration constant to zero: $\Psi_0=0$. This is a key difference from the unrenormalized case, reflecting the fact that holographic renormalization has already ``cleaned'' the UV data. The subleading dilaton coefficients can then be determined iteratively:
\begin{align}\label{Psi1-Psi2-renorm-D}
    \Psi_1 &= \frac{1}{8 \Lambda_0}\left[  \Phi_1 R[q] - \left(\frac{12\pi \ell}{c}\right)^2 \ttbarb \right]\, , \\
    \Psi_2 &= \left(\frac{1}{8 \Lambda_0}\right)^2 \left[ -4 \Lambda_1 + 4 \Lambda_0 \Phi_1 + (2 + \ell^2 \Lambda_0)R[q] \right] \left[  \Phi_1 R[q] - \left(\frac{12\pi \ell}{c}\right)^2 \ttbarb \right]\, .
\end{align}

\paragraph{Resulting effective action.}
The asymptotic expansion of the boundary action to the first non-trivial order is given by:
\begin{equation}\label{JT-induced-Renormalized}
      \mathcal{S}^{\text{\tiny{D}}}_{\text{bdry}}[\Sigma_r] = \mathcal{S}^{\text{\tiny{D}}}_{\text{bdry}}[\Sigma]  + \frac{3\pi \ell^4}{c r^2}\int_{\Sigma} \d{}^{2}x\, \sqrt{-q} \left[ \ttbarb - \left(\frac{c}{12\pi \ell}\right)^2 \Phi_1 R[q] \right] + \mathcal{O}(r^{-4})\, .
\end{equation}
This result is highly significant. In the Fefferman-Graham gauge ($\Phi_1 = 0$), the expansion terminates and becomes exact, successfully recovering the seminal $\text{T}\bar{\text{T}}$ deformation as the unique finite-cutoff modification \cite{McGough:2016lol, Kraus:2018xrn}. However, in our general setup with a non-trivial lapse, we see that the boundary QFT is modified not only by the $\text{T}\bar{\text{T}}$ operator but also by a dynamical coupling between the lapse fluctuation $\Phi_1$ and the boundary curvature—precisely the signature of induced JT gravity. It is also instructive to compare the unrenormalized  \eqref{Sbdry-expansion-check} and renormalized  \eqref{JT-induced-Renormalized} results. The renormalized case, expectedly, does not involve the $r^2$ term and moreover, it does not have the 2d cosmological constant. Note also that the relative coefficients of $\ttbarb$ and $\Phi_1 R[q]$ differ by a factor of two. As a final remark, in order to have a 2d JT gravity with standard conventions one could have defined $\Phi_1$ with a negative sign.

\section{Beyond Dirichlet: A generalized family of boundary conditions}\label{sec:other-bc}
Our analysis so far has focused on the holographic RG flow originating from strict Dirichlet boundary conditions at the AdS boundary. However, the ``GR from RG'' framework is not restricted to this specific choice. To demonstrate the robustness of the RG induced gravity mechanism, we generalize our framework by introducing a one-parameter family of boundary conditions. We show that for this entire family, the radial evolution is universally governed by an induced JT-like dynamics, albeit with shifted effective couplings.

\subsection{Non-renormalized boundary conditions}\label{sec:other-bc-non-R}

\paragraph{A generalized variational principle.}
We consider a family of bulk theories parameterized by a real number $w$. This family is defined by modifying the weight of the Gibbons-Hawking-York term in the bulk action:
\begin{equation}\label{action-AdS-W}
    S^{\text{\tiny{W}}}_{\text{\tiny{bulk}}}[\mathcal{M}_r] = \frac{1}{2\kappa_3} \int_{\mathcal{M}_r} \d{}^3x\, \sqrt{-g} \left(\mathscr{R} + \frac{2}{\ell^2}\right) + \frac{w r^2}{\kappa_3 \ell^2} \int_{\Sigma_r} \d{}^2x\, \sqrt{-\gamma}\, K \, .
\end{equation} 
The parameter $w$ characterizes the specific boundary condition imposed on the cutoff surface. Setting $w=1$ recovers the standard Dirichlet action \eqref{action-AdS}, while other values of $w$ correspond to different mixtures of the metric and its conjugate momentum in the variational principle \cite{Compere:2008us, Adami:2023fbm, Liu:2024ymn, Sheikh-Jabbari:2025kjd, Sheikh-Jabbari:2026ouj, Parvizi:2025shq}.

\paragraph{Generalized deformation flow.}
Following the diffeomorphism-based derivation used in section \ref{sec:deformation-flow-D}, we evaluate the radial evolution of the boundary action $S^{\text{\tiny{W}}}_{\text{bdry}}$. The resulting deformation flow equation takes the form:
\begin{equation}\label{d-f-NR-W-TT}
   \inbox{ \frac{\d{}}{\d{}r} {S}^{\text{\tiny{W}}}_{\text{bdry}}[\Sigma_r] =  - \frac{12 \pi \ell}{c} \left(\frac{\ell}{r}\right)^3 \int_{\Sigma_r} \d{}^2x\, \sqrt{-\gamma}\,  \Phi \qty[ w\,\ttbar +(1-w)\Big(\frac{c}{12 \pi \ell}\Big)^2\frac{2r^4}{\ell^6}]\, .}
\end{equation}
As expected, for $w = 1$, this reduces to the Dirichlet T$\bar{\text{T}}$ flow equation \eqref{d-f-NR-D-TT}. Utilizing the bulk Hamiltonian constraint, we can recast this into its equivalent gravitational formulation:
\begin{equation}\label{d-f-NR-W-Gr}
   \inbox{ \frac{\d{}}{\d{}r} {S}^{\text{\tiny{W}}}_{\text{bdry}}[\Sigma_r] =  \frac{c w}{12\pi r} \int_{\Sigma_r} \d{}^2x\, \sqrt{-\gamma}\,  \Phi \left[ R[\gamma] + \left( 2 - \frac{1}{w} \right)\frac{2r^2}{\ell^4} \right]\, .}
\end{equation}
Remarkably, the flow maintains the structure of the JT gravity action for all $w \neq 0$. The effect of modifying the boundary conditions is entirely captured by a rescaling of the effective Newton's constant ($G_{\text{\tiny{eff}}} \sim 1/w$) and a shift in the effective cosmological constant of the boundary theory.

\paragraph{Integration and the shifted dilaton.}
To integrate this generalized flow, we propose a modified ansatz for the boundary action:
\begin{equation}\label{Ansatz-W}
{S}^{\text{\tiny{W}}}_{\text{bdry}}[\Sigma_r]
= {S}^{\text{\tiny{w}}}_{\text{bdry}}[\Sigma]
- \frac{(2w-1)}{2} \int_{\Sigma_r}\d{}^{2}x\, \sqrt{-\gamma}\,  T
+ \frac{c}{12\pi } \int_{\Sigma_r} \d{}^{2}x\, \sqrt{-\gamma}\,  
\Psi\, \big(R[\gamma] - 2 \Lambda(r) \big)\, .
\end{equation}
Substituting this into the flow equation \eqref{d-f-NR-W-Gr} yields a differential equation for $\Psi$ identical in form to \eqref{Psi-eq}, with the same drag and source coefficients \eqref{X-S-non-D}. The primary distinction lies in the matching conditions at the UV boundary. To ensure the correct asymptotic behavior and maintain consistency with the undeformed theory at infinity, the leading-order dilaton coefficient must be:
\begin{equation}\label{Psi0-w}
    \Psi_0 = - \frac{(2w-1) \Phi_1}{2 \ell^2 \Lambda_0}\, .
\end{equation}
Since all subleading dilaton coefficients $\Psi_{i\geq 1}$ are determined iteratively from $\Psi_0$, they are modified by the boundary condition parameter $w$. This demonstrates that the induced 2d gravity is a universal feature of the AdS$_3$ radial foliation, with the choice of boundary condition merely selecting a specific trajectory within the space of scalar-tensor effective theories.

\paragraph{The critical case $w=1/2$ and conformal boundary conditions.}
An illuminating special case arises when we set $w=1/2$, a choice that famously corresponds to imposing conformal boundary conditions \cite{Anderson:2006lqb, anderson2008boundary, Witten:2018lgb, Coleman:2020jte,  An:2021fcq, Anninos:2023epi, Anninos:2024wpy, Anninos:2024xhc, Liu:2024ymn, Banihashemi:2024yye} (which, uniquely in 3d gravity, also coincides with the Neumann boundary condition \cite{Sheikh-Jabbari:2025kjd, Sheikh-Jabbari:2026ouj}). At this precise value, the effective boundary cosmological constant term proportional to $(2-1/w)$ in the flow equation \eqref{d-f-NR-W-Gr} exactly vanishes. The deformation flow thus elegantly reduces to a purely topological-like curvature coupling:
\begin{equation}
   \frac{\d{}}{\d{}r} {S}^{\text{\tiny{W=1/2}}}_{\text{bdry}}[\Sigma_r] =  \frac{c}{24\pi r} \int_{\Sigma_r} \d{}^2x\, \sqrt{-\gamma}\,  \Phi R[\gamma]\, ,
\end{equation}
and concurrently, the leading-order dilaton coefficient trivializes, $\Psi_0 = 0$. 

Furthermore, the value $w=1/2$ acts as a critical transition point in the parameter space of boundary conditions. As $w$ crosses this threshold, the dilaton seed coefficient $\Psi_0$ \eqref{Psi0-w} undergoes a sign reversal. In the framework of JT gravity, the sign of the dilaton dictates the overall sign of the effective gravitational action; thus, crossing $w=1/2$ formally flips the sign of the boundary metric's kinetic terms. Generally, such a sign change would signal the onset of ghost-like instabilities, highlighting the profound role that the choice of boundary condition plays in securing a well-posed initial-boundary value problem. However, in our specific 3d bulk construction, this does not represent a fatal pathology. Because the radial lapse function enters the bulk line element \eqref{metric} squarely as $\Phi^2$, the bulk geometry enjoys a residual $\mathbb{Z}_2$ symmetry under $\Phi \to -\Phi$ (which equivalently maps $\Phi_i \to -\Phi_i$). One can systematically exploit this gauge freedom to absorb the sign flip, dynamically restoring a strictly positive-definite kinetic term and maintaining a physically healthy induced boundary theory across the threshold.

\subsection{Renormalized boundary conditions}\label{sec:other-bc-R}
Finally, we extend our analysis to a generalized family of \textit{renormalized} boundary conditions \cite{Parvizi:2025shq, Taghiloo:2025oeu}. This family allows us to explore the RG flow of theories that are not purely Dirichlet but involve a specific mixture of the metric and the holographic stress tensor.

\paragraph{Generalized renormalized action.}
We introduce a one-parameter family of bulk theories, parameterized by $w$, defined by the following bulk action:
\begin{equation}\label{action-AdS-R-W}
    \mathcal{S}^{\text{\tiny{W}}}_{\text{\tiny{bulk}}}[\mathcal{M}_r] = \frac{1}{2\kappa_3} \int_{\mathcal{M}_r} \d{}^3x\, \sqrt{-g} \left(\mathscr{R} + \frac{2}{\ell^2}\right) + \frac{r^2}{\ell^2 \kappa_3}\int_{\Sigma_r} \d{}^2x\, \sqrt{-\gamma}\, \left(w\, K - \frac{(2w-1)}{\ell} \right) \, .
\end{equation} 
Crucially, this family possesses a well-defined variational principle for all $w$. Furthermore, unlike the unrenormalized cases discussed in section~\ref{sec:other-bc-non-R}, this construction ensures that the on-shell action and all derived physical quantities remain finite as the regulator is removed ($r \to \infty$). This systematic subtraction of infinite volume divergences is what justifies the designation of this family as ``renormalized'' \cite{Parvizi:2025shq, Taghiloo:2025oeu, Sheikh-Jabbari:2025kjd}.

Invoking the framework of renormalized holography at a finite radial cutoff, the dual boundary action $\mathcal{S}^{\text{\tiny{W}}}_{\text{bdry}}[\Sigma_r]$ satisfies a generalized radial deformation flow equation:
\begin{equation}
    \inbox{\frac{\d{}}{\d{}r} \mathcal{S}^{\text{\tiny{W}}}_{\text{bdry}}[\Sigma_r] = - \int_{\Sigma_r} \d{}^2x\, \sqrt{-\gamma}\, \Phi\, \left( \frac{12 \pi \ell^4}{c\, r^3} w \Ottbar + \frac{\mathcal{T}}{r}  \right)\, .}
\end{equation}
By employing the renormalized Hamiltonian constraint \eqref{EoM-ss-ren}, we recast this flow into its equivalent gravitational formulation:
\begin{equation}\label{flow-eq-grav-renorm-w}
   \inbox{ \frac{\d{}}{\d{}r} \mathcal{S}^{\text{\tiny{W}}}_{\text{bdry}}[\Sigma_r] = \int_{\Sigma_r} \d{}^2x\, \sqrt{-\gamma}\, \Phi\, \left[ \frac{c\, w}{12\pi r} R[\gamma] + (2w-1)\frac{\mathcal{T}}{r}\right]\, .}
\end{equation}

\paragraph{Exact integration and UV limit.}
To integrate the flow \eqref{flow-eq-grav-renorm-w}, we propose an ansatz that includes a holographic trace term to account for the $w$-dependent weighting:
\begin{equation}\label{bdry-action-renorm-w}
\inbox{\mathcal{S}^{\text{\tiny{W}}}_{\text{bdry}}[\Sigma_r]
= \mathcal{S}^{\text{\tiny{W}}}_{\text{bdry}}[\Sigma]
+(1-w)\int_{\Sigma_r} \d{}^2x\, \sqrt{-\gamma}\, \mathcal{T} +  \frac{c}{12\pi} \int_{\Sigma_r} \d{}^{2}x\, \sqrt{-\gamma}\,  
\Psi\, \big(R[\gamma] - 2 \Lambda(r) \big)\, .}
\end{equation}
Substituting this into the flow equation leads back to the same dilaton evolution equation \eqref{Psi-eq-2} derived for the Dirichlet case. Consequently, the integration constant must again be $\Psi_0=0$ to match the undeformed dual theory at infinity. The first non-trivial subleading order of the effective action is then:
\begin{equation}\label{S-bdry-renorm-w-asymp}
      \mathcal{S}^{\text{\tiny{W}}}_{\text{bdry}}[\Sigma_r] = \mathcal{S}^{\text{\tiny{W}}}_{\text{bdry}}[\Sigma]  + \frac{3\pi \ell^4}{c r^2}\int_{\Sigma} \d{}^{2}x\, \sqrt{-q} \left[(2w-1) \ttbarb - \left(\frac{c}{12\pi \ell}\right)^2 \Phi_1 R[q] \right] + \mathcal{O}(r^{-4})\, .
\end{equation}
\paragraph{The non-trivial flow of renormalized conformal boundary conditions.}
The specific choice $w=1/2$ corresponds to imposing the renormalized conformal boundary condition, which  coincides with the renormalized Neumann boundary condition in 3d gravity \cite{Sheikh-Jabbari:2025kjd, Sheikh-Jabbari:2026ouj}. Setting $w=1/2$ yields profound simplifications. In the exact flow equation \eqref{flow-eq-grav-renorm-w}, the explicit dependence on the holographic trace $\mathcal{T}$ perfectly cancels out, leaving a radial evolution driven solely by the curvature term. 

Consequently, in the expansion \eqref{S-bdry-renorm-w-asymp}, the matter-sector $\ttbarb$ deformation completely drops out. If one were to enforce the standard Fefferman-Graham gauge by rigidly fixing the lapse ($\Phi_i = 0$), the expansion formally terminates and the action remains entirely unchanged, yielding exactly $\mathcal{S}^{\text{\tiny{W=1/2}}}_{\text{bdry}}[\Sigma_r] = \mathcal{S}^{\text{\tiny{W=1/2}}}_{\text{bdry}}[\Sigma]$. This perfectly recovers the observation made in \cite{Sheikh-Jabbari:2025kjd, Sheikh-Jabbari:2026ouj} that renormalized conformal/Neumann boundary conditions do not run under the standard holographic RG flow. 

However, our generalized framework reveals a striking new perspective: moving beyond the FG gauge by unfixing the radial lapse demonstrates that this boundary condition \textit{does} in fact flow. Remarkably, this flow is no longer driven by the standard fluidic $\ttbarb$ operator, but is strictly mediated by the emergent JT-gravity geometric deformation $\Phi_1 R[q]$. This encapsulates the essence of the ``GR from RG'' mechanism: the radial lapse activates a purely gravitational RG flow even for boundary conditions that are otherwise completely stable against matter-sector deformations. As we will discuss in section \ref{sec:conc}, this purely geometric flow has a beautiful physical interpretation as the macroscopic manifestation of boundary fluctuations. Whereas, the effects of the boundary condition and $w$ appears in the $\ttbarb$ operator and how it gravitates.

\paragraph{Universality via Legendre transformations.}
An insightful perspective on the ansatz \eqref{bdry-action-renorm-w} emerges through the lens of canonical transformations \cite{Papadimitriou:2010as, Papadimitriou:2004ap, Parvizi:2025shq, Parvizi:2025wsg}. The second term on the right-hand side acts as the generator of a Legendre transformation connecting the Dirichlet and $w$-type boundary conditions at the finite cutoff surface $\Sigma_r$:
\begin{equation}
\mathcal{S}^{\text{\tiny{W}}}_{\text{bdry}}[\Sigma_r] = \mathcal{S}^{\text{\tiny{D}}}_{\text{bdry}}[\Sigma_r] +  (1-w)\int_{\Sigma_r} \d{}^2x\, \sqrt{-\gamma}\, \mathcal{T}\, . 
\end{equation}
Taking the asymptotic limit $r \to \infty$, the Hamiltonian constraint ensures that the integral of $\mathcal{T}$ becomes purely topological (proportional to $\int \sqrt{-q}\, R[q]$). Thus, we establish the equivalence of the on-shell boundary actions at spatial infinity:
\begin{equation}\label{w-D-rel-Sigma}
    \mathcal{S}^{\text{\tiny{W}}}_{\text{bdry}}[\Sigma] = \mathcal{S}^{\text{\tiny{D}}}_{\text{bdry}}[\Sigma] \, .
\end{equation}
This establishes a profound result: the induced JT gravity action we have derived is a universal feature of the AdS$_3$ radial flow. Any theory in the $w$-family can be systematically reconstructed starting from the Dirichlet result and applying the appropriate Legendre transformation. This underscores the fact that the emergence of 2d gravity is an intrinsic geometric property of the radial foliation, independent of the specific boundary condition chosen to anchor the flow.

\section{Discussion and outlook}\label{sec:conc}

In this work, we have provided a concrete and analytical realization of the ``GR from RG'' program \cite{Adami:2025pqr, Sheikh-Jabbari:2026uol} in the context of 2d gravity. By tracking the holographic radial evolution of a generic 2d CFT action, we demonstrated that JT gravity is not merely a phenomenological construct but an inevitable low-energy effective description induced by the renormalization group flow.


\begin{figure}[t]
\centering
\begin{tikzpicture}[scale=1.2]
    
    \begin{scope}[shift={(0,0)}]
        \draw[darkred!50, dashed, thick] (-1.5,0) arc (180:0:1.5 and 0.45);
        
        \shade[left color=red!40!gray!30, right color=red!40!gray!30, middle color=red!5!white, draw=darkred!80, thick, opacity=0.85]
            (-1.5, 0) -- (-1.5, -3.5) arc (180:360:1.5 and 0.45) -- (1.5, 0) arc (360:180:1.5 and 0.45);
            
        \draw[darkred!50, thin, opacity=0.8] (-1.5, -0.7) arc (180:360:1.5 and 0.45);
        \draw[darkred!50, thin, opacity=0.8] (-1.5, -1.4) arc (180:360:1.5 and 0.45);
        \draw[darkred!50, thin, opacity=0.8] (-1.5, -2.1) arc (180:360:1.5 and 0.45);
        \draw[darkred!50, thin, opacity=0.8] (-1.5, -2.8) arc (180:360:1.5 and 0.45);
        
        \foreach \a in {195, 210, 225, 240, 255, 270, 285, 300, 315, 330, 345} {
            \draw[darkred!50, thin, opacity=0.8] ({1.5*cos(\a)}, {0.45*sin(\a)}) -- ({1.5*cos(\a)}, {-3.5 + 0.45*sin(\a)});
        }
        
        \filldraw[fill=gray!20, draw=darkred!80, thick, opacity=0.9] (0,0) ellipse (1.5 and 0.45);
        
        \node[align=center, fill=white, fill opacity=0.9, text opacity=1, rounded corners, inner sep=4pt, draw=darkred!40, drop shadow] at (0, -5.0) {\textbf{Rigid Hypersurface $\Sigma_r$}\\ ($r = \text{const}$ in $r$-frame)\\ Non-trivial lapse $\Phi \neq 1$};
    \end{scope}

    \draw[->, ultra thick, >=stealth, color=black!70] (2.2, -1.75) -- (4.2, -1.75);
    \node[above, font=\small, text=black!80] at (3.2, -1.6) {Gauge Fixing};
    \node[below, font=\small, text=black!80] at (3.2, -1.9) {$r \to \rho(r, x^a)$};

    \begin{scope}[shift={(6.4,0)}]
        
        \draw[darkblue!50, dashed, thick] plot[variable=\t, domain=180:0, samples=80] ({1.5*cos(\t) + 0.05*cos(5*\t)}, {0.45*sin(\t) + 0.04*sin(5*\t)});
        
        \shade[left color=blue!40!gray!30, right color=blue!40!gray!30, middle color=blue!5!white, draw=darkblue, thick, opacity=0.85]
            plot[variable=\y, domain=0:-3.5, samples=60] ({-1.55 - 0.08*sin(-\y*205.714)}, {\y})
            -- 
            plot[variable=\t, domain=180:360, samples=80] ({1.5*cos(\t) + 0.05*cos(5*\t)}, {-3.5 + 0.45*sin(\t) + 0.04*sin(5*\t)})
            --
            plot[variable=\y, domain=-3.5:0, samples=60] ({1.55 + 0.08*sin(-\y*205.714)}, {\y})
            --
            plot[variable=\t, domain=360:180, samples=80] ({1.5*cos(\t) + 0.05*cos(5*\t)}, {0.45*sin(\t) + 0.04*sin(5*\t)})
            -- cycle;
            
        \foreach \h in {-0.7, -1.4, -2.1, -2.8} {
            \draw[darkblue!60, thin, opacity=0.8] plot[variable=\t, domain=180:360, samples=80] 
                ({1.5*cos(\t) + 0.05*cos(5*\t) + cos(\t)*0.08*sin(-\h*205.714)}, {\h + 0.45*sin(\t) + 0.04*sin(5*\t)});
        }
        
        \foreach \a in {195, 210, 225, 240, 255, 270, 285, 300, 315, 330, 345} {
            \draw[darkblue!60, thin, opacity=0.8] plot[variable=\y, domain=0:-3.5, samples=60] 
                ({1.5*cos(\a) + 0.05*cos(5*\a) + cos(\a)*0.08*sin(-\y*205.714)}, {\y + 0.45*sin(\a) + 0.04*sin(5*\a)});
        }
        
        \filldraw[fill=gray!20, draw=darkblue, thick, opacity=0.9] plot[variable=\t, domain=0:360, samples=120] 
            ({1.5*cos(\t) + 0.05*cos(5*\t)}, {0.45*sin(\t) + 0.04*sin(5*\t)});
            
        \node[align=center, fill=white, fill opacity=0.9, text opacity=1, rounded corners, inner sep=4pt, draw=darkblue!40, drop shadow] at (0, -5.0) {\textbf{Wobbly Hypersurface $\Sigma_r$}\\ ($\rho = \rho(x^a)$ in $\rho$-frame)\\ Trivial lapse $\Phi_\rho = 1$};
    \end{scope}

\end{tikzpicture}
\caption{\footnotesize{Geometric interpretation of the induced gravity. \textbf{Left:} In the coordinate system $(r, x^a)$, the physical cutoff $\Sigma_r$ is a constant-radius hypersurface. Here, the non-trivial radial lapse $\Phi \neq 1$ manifests as a dynamical dilaton field. \textbf{Right:} After a radial coordinate transformation to the Fefferman-Graham frame $(\rho, x^a)$, the lapse is fixed to unity ($\Phi_\rho = 1$). However, the physical surface $\Sigma_r$ is no longer at a constant coordinate; it acquires a wobbly profile $\rho(x^a)$, encoding the gravitational reparameterization modes of the boundary.}}
\label{fig:boundary-fluctuations}
\end{figure}
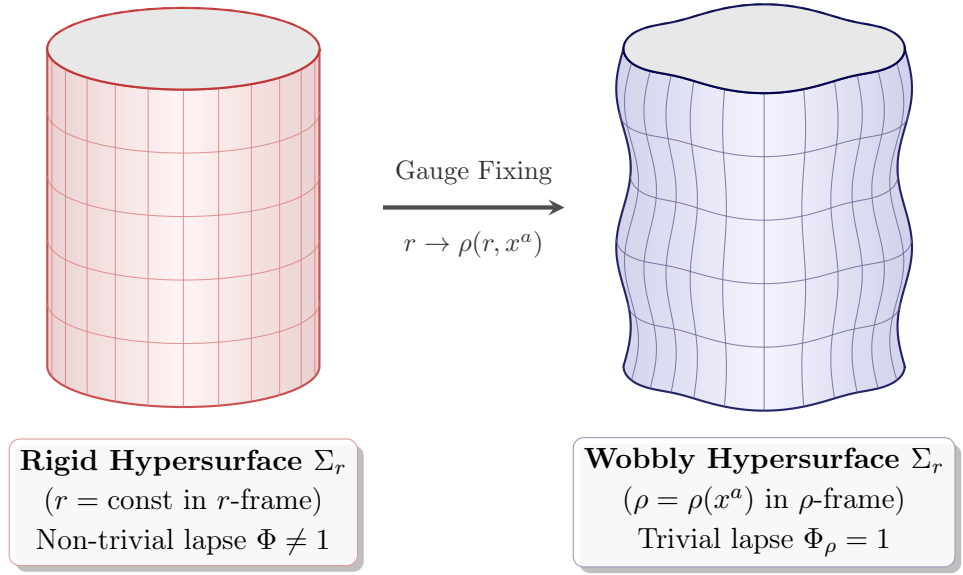


\paragraph{Beyond the Fefferman-Graham gauge.} 
A distinctive feature of our 2d derivation, compared to the higher-dimensional case \cite{Adami:2025pqr}, is the absolute necessity of going beyond the standard FG gauge. In four dimensions, the holographic RG flow induces an Einstein-Hilbert term that is inherently dynamical. In two dimensions, however, the Einstein-Hilbert action is purely topological. If one strictly adheres to the FG gauge ($\Phi=1$), the induced action at the cutoff surface remains a topological invariant, failing to capture the rich gravitational dynamics characteristic of the IR. Our results show that the ``activation'' of 2d gravity is explicitly tied to the radial lapse function $\Phi$. By retaining $\Phi$ as a generic degree of freedom, the topological term $\int R$ is upgraded to the dynamical JT term $\int \Psi R$, where the lapse effectively seeds the dilaton.

\paragraph{Induced gravity as boundary fluctuations.}
This leads to a deep geometric and physical question: since the lapse function $\Phi$ can be fixed to unity by a bulk coordinate transformation, what is the physical origin of this induced gravity? To see this clearly, consider our starting point: a coordinate system $(r, x^a)$ where the radial lapse $\Phi(r, x^a) \neq 1$ and the physical cutoff $\Sigma_r$ is defined as a surface of constant radius $r$. 

One can always perform a radial coordinate transformation $r \to \rho(r, x^a)$ to reach a new frame $(\rho, x^a)$ where the lapse is set to unity ($\Phi_\rho = 1$), corresponding to the standard FG gauge. However, this transformation redefines the meaning of the radial scale. Crucially, the original physical cutoff $\Sigma_r$ (which was a uniform, constant-coordinate hypersurface in the $r$-frame) is no longer a surface of constant coordinate in the new $\rho$-frame. Instead, as illustrated in Figure \ref{fig:boundary-fluctuations}, the physical surface $\Sigma_r$ is mapped to a non-trivial, ``wobbly'' profile $\rho = f(x^a)$. 

This provides a beautiful and unified picture of emergent gravity in 2d: \textit{induced gravity is a manifestation of the fluctuations of the holographic cutoff surface.} In the rigid $(r, x^a)$ frame, the boundary is straight, and the gravitational degrees of freedom appear as a dynamical dilaton field $\Psi$ sourcing the curvature. In the FG $(\rho, x^a)$ frame, the lapse is trivial, but the gravitational dynamics are now encoded in the reparameterization modes (the ``wobble'') of the physical cutoff surface relative to the FG scale \cite{Maldacena:2016upp, Jensen:2016pah, Engelsoy:2016xyb}. Our derivation confirms that the induced JT gravity is not an auxiliary structure, but an intrinsic description of the boundary’s capacity to fluctuate—a feature that is made manifest by the bulk radial lapse under holographic RG flow.

\paragraph{Unfreezing the metric.}
A persistent conceptual hurdle in the emergent gravity paradigm is the question of how a fixed background metric in the UV can become a dynamical, fluctuating field in the IR. We have addressed this through the \textit{RG flow of boundary conditions} \cite{Adami:2025pqr, Sheikh-Jabbari:2026uol}. We showed that a rigid Dirichlet boundary condition imposed at asymptotic infinity ($\delta \gamma_{ab} = 0$) does not remain Dirichlet at finite scales. Instead, it systematically evolves into a mixed Dirichlet-Neumann condition, where the metric fluctuations are functionally tied to the energy-momentum tensor. This ``unfreezing'' mechanism suggests that the transition from a non-gravitating QFT to a gravitating theory happens as a result of integrating  out UV degrees of freedom, and that the irrelevant T$\bar{\text{T}}$ deformation can be recast as boundary gravity.

\paragraph{Dirichlet-to-Neumann flow as the genesis of gravity.} 
An intriguing feature of this construction is the natural transition from a Dirichlet boundary condition in the UV to a Neumann condition in the IR. As explored in section \ref{sec:Neumann-interp}, the exact integration of the deformation flow inherently provides the Legendre transformation needed to redefine the variational principle at the cutoff surface. This transition plays a functional role in the emergence of gravity: by reaching a Neumann condition, the induced metric is no longer tethered to a fixed background structure. Instead, it becomes a dynamically varying field, which is a fundamental requirement for any theory of gravity. In this sense, the holographic RG flow does more than just induce the gravitational action; it establishes the appropriate variational framework for the metric to function as a dynamical degree of freedom.

In conclusion, the ``GR from RG'' program suggests that the search for a quantum theory of gravity by quantizing the metric may be fundamentally misplaced. If the metric and its dynamics are collective IR descriptions of QFT degrees of freedom, the graviton is a collective excitation akin to a phonon in a fluid. Our derivation of JT gravity from the 2d CFT lapse provides a robust, lower-dimensional laboratory for this emergent viewpoint, suggesting that gravity is not a fundamental force, but rather the hydrodynamics of quantum field theory under coarse-graining.

\section*{Acknowledgement}
We thank Mahdi Golshani and Mohammad Hassan Vahidinia for their contributions in the early stages of this work. VT would also like to thank Vahid Reza Shajiee for useful discussions. We also acknowledge the stimulating scientific atmosphere of the Simons Center workshop \textit{Timelike Boundaries in Classical and Quantum Gravity} (December 2025). The work of VT is supported by the Iran National Science Foundation (INSF) under project No. 4040771. The work of MMShJ and VT is supported in part by the INSF Research Chair grant No. 4045163.

\appendix
\section{Useful relations}\label{sec:relations}
In this appendix, we collect several identities and radial evolution equations for the geometric and boundary variables. These relations are essential for verifying the consistency of the holographic RG flow and for performing the exact integration of the boundary action.

\paragraph{Evolution of metric components.}
Using the radial foliation and the definition of the extrinsic curvature \eqref{Kab}, the radial evolution of the determinant and the inverse induced conformal metric is given by:
\begin{equation}\label{dr-h-1}
\begin{split}
    \mathcal{D}_r \sqrt{-\gamma} & = -\frac{2}{r}\sqrt{-\gamma} - \frac{\kappa_3 \ell^3}{r^3} \sqrt{-\gamma}\, \Phi\,  T\, ,\\
    \mathcal{D}_r \gamma^{ab} & = \frac{2}{r} \gamma^{ab} - \frac{2\kappa_3 \ell^3}{r^3} \Phi (T^{ab} - T \gamma^{ab})\, .
\end{split}
\end{equation}

\paragraph{Stress tensor and curvature dynamics.}
Combining the bulk Einstein equations with the identities above, we derive the radial evolution for the boundary-weighted stress tensor and curvature scalars:
\begin{subequations}\label{EoM-TR-r}
\begin{align}
    \mathcal{D}_r (\sqrt{-\gamma}\, T^{ab}) & = \sqrt{-\gamma} \left[ \frac{2}{r}T^{ab} + \frac{\ell \Phi}{2r \kappa_3} R \gamma^{ab} - \frac{\ell}{r \kappa_3} (\nabla^{a}\nabla^{b}\Phi - \gamma^{ab} \Box \Phi) \right]\, , \label{dr-T-up}\\
    \mathcal{D}_r{(\sqrt{-\gamma}\, T)} & = -\sqrt{-\gamma}\frac{\ell\Phi}{r\kappa_3}\left(R + \frac{4r^2}{\ell^4} - \frac{\Box{\Phi}}{\Phi}\right) \,,\\
     \mathcal{D}_r{{T}} & = \frac{2T}{r} - \frac{\ell \Phi}{ r \kappa_3} \left[ R + \frac{4r^2}{\ell^2} - \left( \frac{\ell \kappa_3 T}{r} \right)^2 - \frac{\Box \Phi}{\Phi}\right]\, , \\
    \mathcal{D}_r (\sqrt{-\gamma}\, R) & = \frac{2\ell^3\kappa_3}{r^3} \sqrt{-\gamma}\,   T^{ab}\nabla_{a}\nabla_b \Phi = \frac{2\ell^3\kappa_3}{r^3} \nabla_{a}\left(\sqrt{-\gamma}\,   T^{ab}\nabla_b \Phi\right)\, .
\end{align}
\end{subequations} 

\paragraph{Evolution of the JT-block.}
The specific combination of terms that constitutes the core of the induced JT gravity action evolves as:
\begin{equation}
    \mathcal{D}_r \left[\sqrt{-\gamma} \left(R+\frac{2r^2}{\ell^4} \right) \right] = 2 \kappa_3\sqrt{-\gamma}  \left( -\frac{\Phi T}{\ell r} + \frac{\ell^3}{r^3} T^{ab} \nabla_a \Phi \nabla_b \Phi \right)\, .
\end{equation}


\bibliographystyle{fullsort.bst}
\bibliography{reference}


	\end{document}